\documentclass[fleqn,usenatbib]{mnras}
\usepackage{comment}
\usepackage{gensymb}
\usepackage{refcount}
\usepackage{tabularx}
\usepackage{enumitem}
\usepackage{adjustbox}
\usepackage{multirow}
\usepackage{pifont}
\usepackage{siunitx}

\usepackage[T1]{fontenc}
\usepackage{ae,aecompl}
\usepackage{xcolor}
\usepackage{threeparttable}

\newcommand{\HI}{H\,{\sc i}}

\usepackage{graphicx}	
\usepackage{savesym}
\usepackage{amsmath}
\savesymbol{iint}
\usepackage{txfonts}
\restoresymbol{TXF}{iint}
\makeatletter
\newcommand\footnoteref[1]{\protected@xdef\@thefnmark{\ref{#1}}\@footnotemark}
\def\env@cases{%
  \let\@ifnextchar\new@ifnextchar
  \left\lbrace
  \def\arraystretch{1.2}%
  \array{l@{\quad}l@{}}
}
\makeatother

\title{The extended \HI\ halo of NGC~4945 as seen by MeerKAT}
\author[Ianjamasimanana R. et al.]
{Roger Ianjamasimanana$^{1,2,3}$\thanks{E-mail: ianja@iaa.es; ianja@starscientist.org},
B.\ S.\ Koribalski,$^{4,5}$
Gyula I. G. J{\'o}zsa,$^{6,2,3}$
Peter Kamphuis,$^{7}$ \newauthor
W. J. G. de Blok,$^{8,9,10}$
Dane Kleiner,$^{11}$
Brenda Namumba,$^{2}$
Claude Carignan,$^{12, 13, 14}$ \newauthor
Ralf-J\"{u}rgen Dettmar,$^{7}$ 
Paolo Serra,$^{11}$
Oleg M. Smirnov,$^{2,3}$
Kshitij Thorat,$^{15, 16}$ \newauthor
Benjamin V. Hugo,$^{2,3}$ 
Athanaseus J. T. Ramaila,$^{3}$ 
Eric Maina,$^{2}$
Filippo M. Maccagni,$^{11}$ \newauthor
Sphesihle Makhathini,$^{17}$ 
Lexy A. L. Andati,$^{2}$ 
D\'{a}niel Cs. Moln\'{a}r,$^{11}$ 
Simon Perkins,$^{3}$ \newauthor
Francesca Loi$^{11}$, Mpati Ramatsoku,$^{2,11}$
Marcellin Atemkeng$^{2}$
\\
$^{1}$Instituto de Astrofísica de Andalucía (CSIC), Glorieta de la Astronomía, E-18008 Granada, Spain\\
$^{2}$Department of Physics and Electronics, Rhodes University, PO Box 94, Makhanda, 6140, South Africa\\
$^{3}$South African Radio Astronomy Observatory, 2 Fir Street, Black River Park, Observatory, Cape Town, 7925, South Africa\\
$^{4}$Australia Telescope National Facility, CSIRO Astronomy and Space Science, P.O. Box 76, NSW 1710, Epping, Australia \\
$^{5}$Western Sydney University, Locked Bag 1797, Penrith, NSW 2751, Australia\\
$^{6}$Max-Planck-Institut f\"ur Radioastronomie, Radioobservatorium Effelsberg, Max-Planck-Stra{\ss}e 28, 53902 Bad M\"unstereifel, Germany\\
$^{7}$Ruhr University Bochum, Faculty of Physics and Astronomy, Astronomical Institute, 44780 Bochum, Germany\\
$^{8}$Netherlands Institute for Radio Astronomy (ASTRON), Postbus 2, 7990 AA Dwingeloo, the Netherlands\\
$^{9}$Dept.\ of Astronomy, Univ.\ of Cape Town, Private Bag X3, Rondebosch 7701, South Africa\\
$^{10}$Kapteyn Astronomical Institute, University of Groningen, Postbus 800, 9700 AV Groningen, The Netherlands\\
$^{11}$INAF - Osservatorio Astronomico di Cagliari, Via della Scienza 5, I-09047 Selargius (CA), Italy\\
$^{12}$Department of Astronomy, University of Cape Town, Private Bag X3, Rondebosch 7701, South Africa\\
$^{13}$D\'{e}partement de physique, Universit\'{e} de Montr\'{e}al, Complexe des sciences MIL, 1375 Avenue \\ \hspace{0.1cm} Th\'{e}r\`{e}se-Lavoie-Roux Montr\'{e}al, Qc, Canada H2V 0B3 \\
$^{14}$Laboratoire de Physique et de Chimie de l'Environnement, Observatoire d'Astrophysique de l'Universit\'{e} Ouaga I \\ \hspace{0.1cm} Pr Joseph Ki-Zerbo (ODAUO), BP 7021, Ouaga 03, Burkina Faso\\
$^{15}$Department of Physics, University of Pretoria, Hatfield, Pretoria, 0028, South Africa\\
$^{16}$The Inter-University Institute for Data Intensive Astronomy (IDIA), Department of Astronomy, University of Cape Town, \\ \hspace{0.1cm} Private Bag X3, Rondebosch, 7701, South Africa\\
$^{17}$University of the Witswatersrand, 1 Jan Smuts Avenue, Braamfontein 2000, Johannesburg, South Africa\\
}

\date{Accepted 2022 March 31. Received 2022 March 10; in original form 2021 March 18}

\pubyear{2022}

\begin{document}
\label{firstpage}
\pagerange{\pageref{firstpage}--\pageref{lastpage}}
\maketitle
\begin{abstract}
Observations of the neutral atomic hydrogen (\HI) in the nuclear starburst galaxy NGC~4945 
with MeerKAT are presented. We find a large amount of halo gas, previously missed by \HI\ observations, accounting for 6.8\% of the total \HI\ mass. 
This is most likely gas blown into the halo by star formation. Our maps go down to a $3\sigma$ column density level of \num{5e+18} $\mathrm{cm^{-2}}$. 
We model the \HI\ distribution using tilted-ring fitting techniques and find a warp on the galaxy's approaching and receding sides. The \HI\ in the northern side of the galaxy appears to be suppressed. 
This may be the result of ionisation by the starburst activity in the galaxy, as suggested by a previous study. The origin of the warp is unclear but could be due to past interactions or ram pressure stripping.
Broad, asymmetric \HI\ absorption lines extending throughout the \HI\ emission velocity channels are present towards the nuclear region of NGC~4945. 
Such broad lines suggest the existence of a nuclear ring moving at a high circular velocity. 
This is supported by the clear rotation patterns in the \HI\ absorption velocity field. 
The asymmetry of the absorption spectra can be caused by outflows or inflows of gas in the nuclear region of NGC~4945. The continuum map shows small extensions 
on both sides of the galaxy's major axis that might be signs of outflows resulting from the starburst activity. 
\end{abstract}




\section{Introduction}
\indent Starburst galaxies are characterised by their intense burst of star formation over a short period \citep{1996A&A...309..345C}. 
Their star formation rates range from $\sim10~M_{\odot}~yr^{-1}$ to $\sim1000~M_{\odot}~yr^{-1}$. 
There are sub-classes of starburst galaxies; those with a vigorous nuclear star formation, e.g., NGC~4945, NGC 253, NGC 660, NGC 1068, NGC 1365, NGC 1808, NGC 3079 
\citep{1996ASPC..106..238K, 2020ApJ...903...50E}
and those that have a more distributed star formation mode, e.g., NGC~4631 \citep{2011MNRAS.410.1423I, 2012AJ....144...44I}. 
This paper focuses on the study of the edge-on nuclear starburst galaxy NGC~4945. 
For simplicity, we will refer to it simply as a starburst galaxy throughout the paper.  
The high levels of star formation in the nuclear region of starburst galaxies are thought to be due to gas accretion onto the centre, 
induced by bars, galaxy interactions or 
past merging events \citep{1990Natur.344..224I, 2001A&A...372..463O, 2015MNRAS.450.3935L}. Starburst galaxies have 
stellar populations dominated by massive stars, which transfer a significant amount of energy into the interstellar medium 
and, as a result, shape the gas distribution and kinematics of the galaxies \citep{Freyer_2003, 2018Natur.558..260Z}. Simulations show that star formation can 
drive the gas out of the disc plane, mostly as 
ionized gas. The gas eventually cools and rains back to the disc, fuelling future episodes of star formation. 
This so-called \textit{galactic fountain model} \citep{1976ApJ...205..762S} is consistent with observations of starburst galaxies.
For example, extra-planar gas has been observed in the starburst galaxy NGC~4945 \citep{2018agn..confE..50V} and its twin NGC~253 
\citep{2005A&A...431...65B, 2015MNRAS.450.3935L}, 
as well as the dwarf starburst galaxy M\,82 \citep{2018ApJ...856...61M}. Edge-on galaxies offer a better way to study the vertical distribution of the 
gas compared to galaxies seen 
more face-on. For example, deep Westerbork Radio Telescope (WSRT) \HI\ 
observations of the edge-on galaxy NGC~891 by \citet{2007AJ....134.1019O}, labelled as a quiescent starburst galaxy by \citet{2005MNRAS.362..581T}, 
show large quantities of extra-planar gas, extending up to a vertical distance of 22 kpc. Thus, observations of the gas content of
starburst galaxies are important in the studies of the mechanisms that
initiate and regulate starburst activity. \\\\
\indent The neutral atomic hydrogen gas (H\,{\sc i}) remains the best tracer of the overall gas distribution and 
kinematics in galaxies. This is because H\,{\sc i} can be traced out to a considerable distance from the galactic centre. 
Previous H\,{\sc i} 
observations of nearby starburst galaxies have allowed 
the investigations of the possible triggering mechanisms of starburst. For example, the early H\,{\sc i} observations of M\,82  
revealed that its velocity field showed strong deviations from circular motions, suggesting possible interactions with M81. Subsequent H\,{\sc i} observations 
show filamentary structures connecting M\,81, M\,82 and NGC~3077, forming what is 
called the M\,81-triplet \citep{1994Natur.372..530Y, 2018ApJ...865...26D, 2019MNRAS.486..504S}. 
The interaction of M\,82 with its neighbour is thought to have triggered its starburst activity \citep{2006ApJ...649..172M}. 
H\,{\sc i} observations of NGC~253 by \citet{2015MNRAS.450.3935L} 
unveiled extra-planar gas emission out to a projected distance of $\sim~9$ to $\sim~10$ kpc from the nucleus and up to 14 kpc at the edge of the disc. 
Extra-planar gas has also been found for other starburst galaxies such as NGC 3079, NGC~4945, NGC~891 
\citep{1997ApJ...491..140S, 2001A&A...372..463O, 2001ApJ...562L..47F, 2005A&A...431...65B, 2007AJ....134.1019O, 2013A&A...554A.125G, 2019A&A...631A..50M}. 
Extra-planar gas is also common in normal star-forming spiral galaxies. For example, The Hydrogen Accretion in LOcal GAlaxieS survey \citep[HALOGAS,][]{2011A&A...526A.118H} found halo gas 
enclosing about 5--25\% of the total \HI\ mass \citep{2019A&A...631A..50M}, and moving slower than the gas in the disc of normal galaxies.  \\[.2in] 
\indent The amount of extra-planar H\,{\sc i} found in starburst galaxies by previous observations is within the range found 
for normal galaxies \citep{2005A&A...431...65B, 2015MNRAS.450.3935L}.
With the improved sensitivity of MeerKAT, we aim to get high-quality H\,{\sc i} imaging to trace possible halo gas that might have been missed by previous H\,{\sc i} observations. 
In addition, we would like to compare the amount and distribution of the extra-planar H\,{\sc i} we found for our starburst galaxy with those for normal galaxies. 
In this paper, we report high sensitivity H\,{\sc i} observations of the nuclear starburst galaxy NGC~4945 with MeerKAT. Our H\,{\sc i} map is about 2.5 times deeper 
in column density than previous ATCA mosaic imaging by LVHIS \citep{2018MNRAS.478.1611K}. NGC~4945 is a nearby edge-on starburst galaxy 
belonging to the Centaurus A group. The distance to NGC~4945 quoted in the literature varies between 3.8 to 8.1 Mpc 
\citep{1964ApJ...139..899D, 1985Natur.315...26B, 1992A&A...265..403B, 2001A&A...372..463O, 2007AJ....133..504K, 2008ApJ...686L..75M}. 
Here we adopt a distance of 6.7 Mpc for easy comparison with the results of \citet{2001A&A...372..463O} and other relevant studies. 
NGC~4945 is classified as a SB(s)cd galaxy \citep{1991rc3..book.....D}. 
The central region of NGC~4945 harbours several physical processes. Apart from hosting a nuclear starburst, 
NGC~4945 is known to have a Seyfert\,2 nuclei, which is thought to be responsible for some of its out-flowing gas \citep{1996ApJ...463L..63D}. 
Moreover, the nuclear region of NGC~4945 is among the richest sources of molecular lines, and therefore it has extensively 
been studied in the millimetre wavelengths 
\citep{1993A&A...270...29D, 1994A&A...284...17H, 2001A&A...367..457C, 2001A&A...372..463O, 2008A&A...479...75H, 2016ApJ...827...69H, 2018A&A...615A.155H}. 
Measurements of different molecular abundance ratios in NGC~4945 suggest that its starburst activity is possibly reaching an advanced evolutionary 
stage \citep{2001A&A...367..457C, 2005MNRAS.364...37C}. Some authors classified NGC~4945 as a post-starburst galaxy \citep{1993ApJ...403..581K}. 
However, this was not confirmed by subsequent studies. The H\,{\sc i} disc of NGC~4945 radially extends just a bit larger than the bright optical 
disc as shown by the ATCA map of \citet{2018MNRAS.478.1611K} and our MeerKAT data shown in this paper. 
This is peculiar for late-type spiral galaxies whose H\,{\sc i} disc usually extends well beyond the optical disc \citep{2001A&A...372..463O, 2018MNRAS.478.1611K}. 
Other starburst galaxies that share this peculiarity include NGC~253 \citep{2018MNRAS.478.1611K} and NGC~1433 \citep{1996ApJ...460..665R}.\\[.2in]
\indent In this paper, we show high sensitivity maps of the radio continuum emission and the H\,{\sc i} spectral 
line emission/absorption in NGC~4945 obtained with MeerKAT. 
Aided by the large field of view of MeerKAT, we obtained, for the first time, the most complete H\,{\sc i} map of NGC~4945 with only a single pointing. 
This allows us to trace potential star formation outflows much more easily \citep{2015MNRAS.450.3935L, 1996ApJ...460..665R}.

\section{Observations and data reduction}
NGC~4945 was observed with MeerKAT during 24--26 of May 2019. The data was taken over two observing epochs using the L-band receiver, centred at 1284 MHz, 
with a bandwidth of 856 MHz divided into 4096 channels (44 $\mathrm{km~s^{-1}}$ in velocity resolution). The compact radio sources PKS 1934--63, PKS 0408--65, 
and J1331+3030 were used as the bandpass calibrators, and J1318--4620 as the gain calibrator. 
Each observing run starts and ends with the bandpass calibrator, which was tracked every two hours. The target and the gain calibrator were tracked one after the other. 
Thus, we spend 10 mins on bandpass every 2 hours, within which we cycle through 15 mins target exposure and 2 mins gain calibrator observation. 
This results in a total observing time of 15 hours 50 minutes, during which 12 hours 50 minutes were spent on the target.\\

The CARACal\footnote{https://github.com/caracal-pipeline/caracal} \citep{2020ascl.soft06014J} data-reduction pipeline was used to produce a 
spectral-line data cube and a continuum image ready for scientific analysis. Based on  Python/Stimela\footnote{https://github.com/ratt-ru/Stimela/wiki} scripts, 
CARACal incorporates many tasks from standard data-reduction software, e.g., CASA \citep{2007ASPC..376..127M}, MeqTrees \citep{MeqTrees}, 
CubiCal \citep{2018MNRAS.478.2399K}, 
SoFiA \citep{2015MNRAS.448.1922S}, 
which can be run sequentially using a YAML (a human-readable data-serialisation language) configuration file through which the users can 
control which tasks need to be run/turned off and in which order. 
CARACal can perform all the standard data-reduction steps of radio astronomical data, from flagging, calibration to imaging. 
CARACal is written in Python and uses container-based data processing technologies such as Singularity, Docker, uDocker or Podman, 
allowing it to run across different platforms. Although CARACal was initially developed to reduce data from MeerKAT, its scope has been 
broadened to allow data processing from any other radio telescopes. Examples of peer-reviewed papers using CARACal include \citet{2020MNRAS.497.4795I}, 
\citet{2020A&A...643A.147D}, \citet{2021MNRAS.501.2704J}. \\\\
\indent We use a bandwidth of 30 MHz, centred around the \HI\ line of NGC 4945. To produce the final data used in this analysis, first, radio frequency interference (RFI) were 
flagged using the AOFlagger \citep{2012PhDT........23O}. 
Then, cross-calibration was performed using the CASA task \textit{gaincal} to solve for 
the time-dependent delays (K) and complex gains (G) of each antenna, 
and the \textit{bandpass} task to obtain the bandpass (B) solutions. 
After that, the data were passed through the CARACal selfcal worker to perform self-calibration. 
For that, five successive WSClean \citep{2014MNRAS.444..606O} imaging loops and phase-only gain corrections 
with CubiCal \citep{2018MNRAS.478.2399K} were used. The clean mask threshold was iteratively lowered in each step until 
the continuum residual was noise-like. Finally, to produce the spectral line data cube, first, the continuum clean model produced in the self-calibration step was   
subtracted from the calibrated data. After that, the CASA task \textit{mstransform} was used to fit the remaining continuum 
with a polynomial function of order one and subtract it from the data. The continuum subtracted data were then cleaned using WSClean in its multi-scale 
mode. A robust parameter of 0 was used, which resulted in a synthesised beam size of 
\SI{7.5}{\arcsecond} $\times$ \SI{6.4}{\arcsecond}. The rms noise of the final data cube is 0.085 $\mathrm{mJy~beam^{-1}}$, which is close to the expected theoretical 
noise of 0.070 $\mathrm{mJy~beam^{-1}}$, calculated from the radiometer formula and using a penalty factor 
of 1.5 to account for robust weighting. We further smooth the data to a \SI{60}{\arcsecond} 
circular beam with the MIRIAD~\citep{1995ASPC...77..433S} 
task CONVOL. The rms noise of the \SI{60}{\arcsecond} cube is 0.20 $\mathrm{mJy~beam^{-1}}$. 
Note that the detected halo gas reported in this paper is already clearly visible even at the highest spatial resolution 
(i.e., \SI{7.5}{\arcsecond} $\times$ \SI{6.4}{\arcsecond}) as shown in the Appendix. 
One may think that smoothing the cube to a larger beam after cleaning could result in large-scale wings and ripples in the dirty psf mimicking halo gas. 
However, our use of a multi-scale cleaning, which uses different size scales, already addresses these issues. In addition, we have also 
used a mask produced by SoFiA during an initial imaging step to produce our final image. 
We show the low-resolution version in this paper to emphasise the halo gas and speed up the processing 
time for the 3D kinematic modelling. Thus, the results presented in the main section of the paper are those using the smoothed data cube unless stated 
otherwise in the paper. In addition, we made two versions of the radio continuum map of NGC 4945, one at a resolution of \SI{17.6}{\arcsecond} $\times$ \SI{15.8}{\arcsecond}, and 
another one at a resolution of \SI{7.3}{\arcsecond} $\times$ \SI{6.2}{\arcsecond}. This paper presents the continuum map at 
\SI{17.6}{\arcsecond} $\times$ \SI{15.8}{\arcsecond} unless otherwise stated.

\section{Results}

\begin{figure*} 
\centering
\includegraphics[scale=0.66]{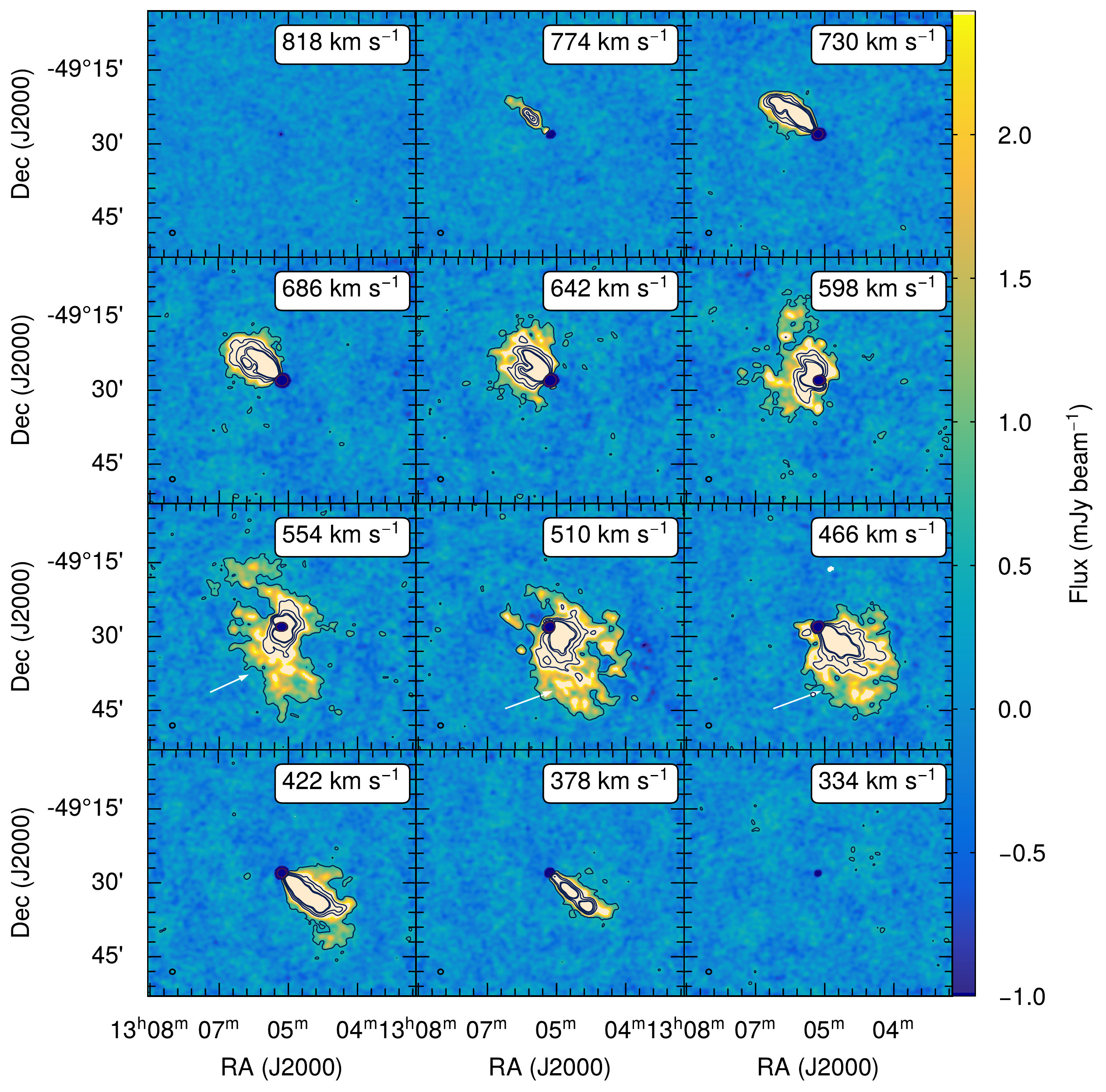}
\caption{MeerKAT H\,{\sc i} channel maps of the nearby starburst galaxy NGC~4945, smoothed to a resolution of \SI{60}{\arcsecond}. 
The contour levels are (--10, --50, --90, --130, --170, --110, --150, 3, 20, 40, 80, 100) $\times$ rms where the rms is 0.2 mJy\,beam$^{-1}$. 
The red contours show the absorption in the centre of the galaxy. The white arrows indicate some of the previously unseen halo gas mentioned in the text.
The black circles at the bottom left corner of each panel show the size of the beam  (FWHM = \SI{60}{\arcsecond}).}
\label{fig:NGC4945-ch}
\end{figure*}

\subsection{\HI\ channel maps and global profile}
\begin{figure} 
\begin{tabular}{l}
\includegraphics[scale=0.55]{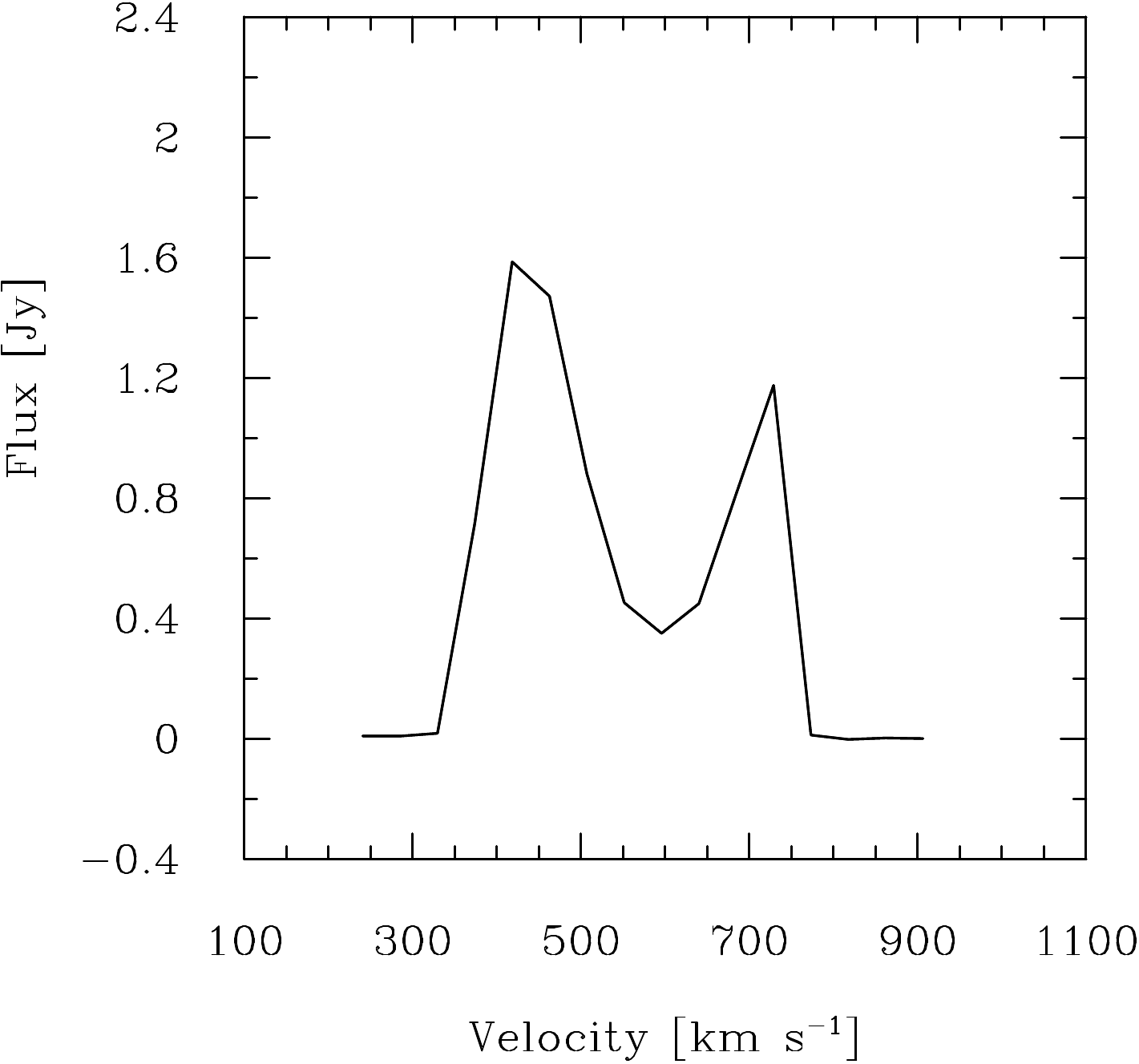}
\end{tabular}
\caption{MeerKAT Global H\,{\sc i} profile of NGC~4945.} 
\label{fig:NGC4945-gl}
\end{figure}

We present the MeerKAT H\,{\sc i} channel maps of the starburst galaxy NGC~4945 in Fig.~\ref{fig:NGC4945-ch}. 
They reveal previously unseen fluffy faint H\,{\sc i} emission surrounding and/or connected to the main disc of the galaxy. 
They are most clearly visible at channels $V$ = (466, 510, 554, 598) km\,s$^{-1}$ as indicated by the white arrows in the Figure. 
We are unsure about their exact origin due to the lack of velocity resolution. 
We could not model them properly and have excluded them in our modelling presented in Section~\ref{sec:modelling}. 
We will discuss its possible origin in Section~\ref{sec:discussion}. We present the global H\,{\sc i} profile of NGC~4945 in Fig.~\ref{fig:NGC4945-gl}. 
The global profile is double-horned, typical of spiral galaxies with a flat rotation curve. 
In addition, it is asymmetric, with the approaching side having higher peak flux-density than the receding side.
By fitting the profile using the busy function \citep[BusyFit,][]{2014MNRAS.438.1176W}, we find a $W_{50}$ 
(full width at 50\% of the peak flux) of 359 km~s$^{-1}$, a $W_{20}$ (full width at 20\% of the peak flux) of 397 km~s$^{-1}$, and a systemic velocity of 
$v_{sys}$ = 530 km~s$^{-1}$.  Note that the $v_{sys}$ we use throughout the rest of the paper is the one we estimated from 
the tilted-ring modelling described in section~\ref{sec:modelling}. 
From the \HI\ intensity map we measure a total \HI\ flux of 509 Jy\,km~s$^{-1}$. Single dish observations 
using the Parkes radio telescope resulted in a measured flux of 319 Jy\,km~s$^{-1}$, 
$v_{sys}$ = 563 km~s$^{-1}$, and $W_{50}$ = 361 km~s$^{-1}$ \citep[HIPASS,][]{2004AJ....128...16K}. The ATCA mosaic imaging by \citet{2018MNRAS.478.1611K} yielded a total flux of 405.3 Jy\,km~s$^{-1}$. 
Thus, we detect 37\% and 20\% more flux than the HIPASS and the LVHIS surveys, respectively. Using our adopted distance of 6.7 Mpc, 
we measured a total \HI\ mass of \num{5.4e+9} $M_{\odot}$.    
Recall that here only a single pointing was required to map the extended H\,{\sc i} disc of NGC~4945, which is thanks to the large primary 
beam of MeerKAT. Due to the strong \HI\ absorption present towards the nuclear region of the galaxy, 
the estimated total mass represents the H\,{\sc i} mass 
seen in emission but not the total H\,{\sc i} mass. 


\subsection{\HI\ moment maps}\label{sub:moment} 
\begin{figure*} 
\begin{tabular}{l} 
\hspace{-0.5cm} \includegraphics[scale=0.62]{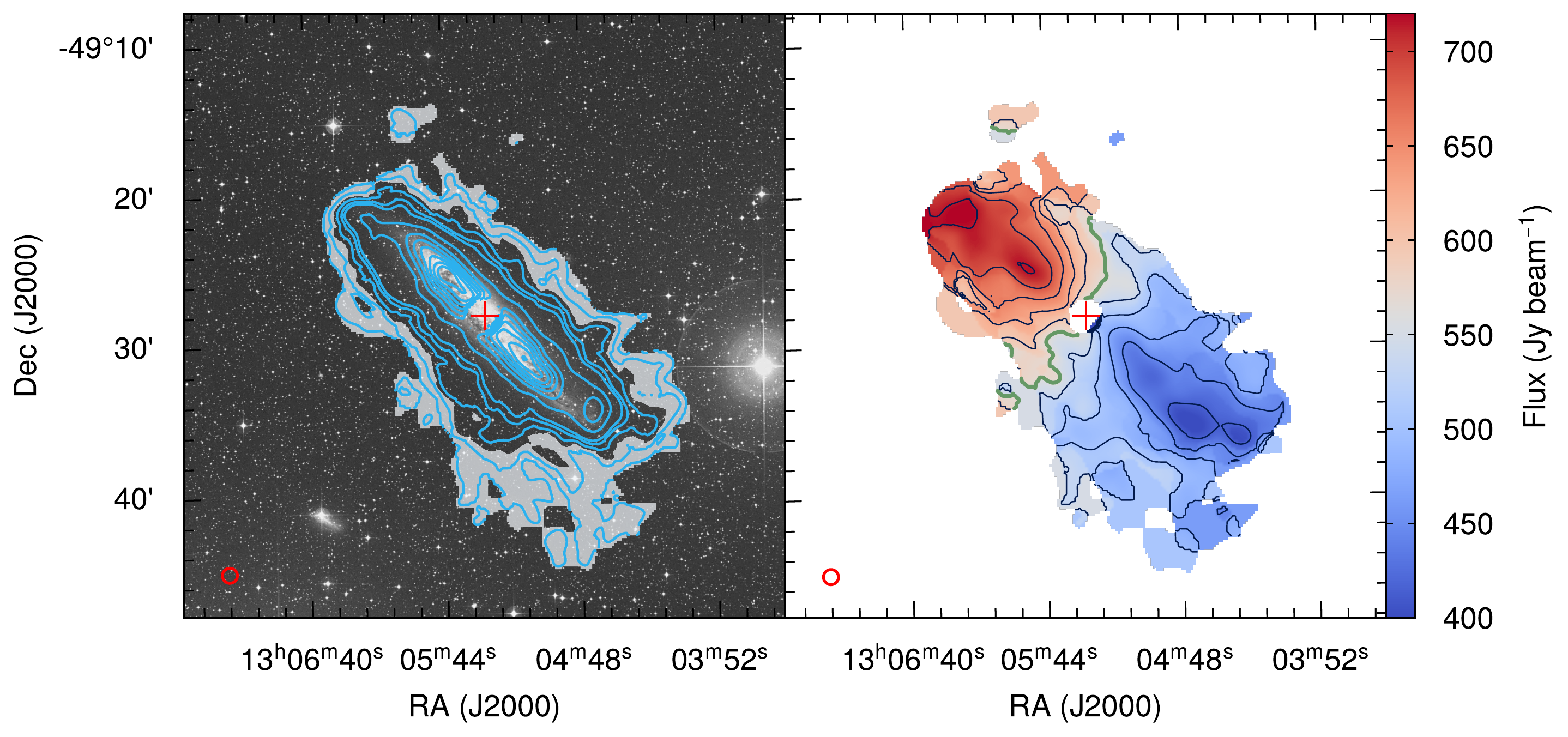} 
\end{tabular}
\caption{Left; MeerKAT \HI\ column density map of the starburst galaxy 
NGC~4945 overlaid onto a DSS2  $B$-band optical image shown in grayscale. 
The contour levels are (\num{4.1e+19}, \num{1.0e+20}, \num{2.0e+20}, \num{3.1e+20}, \num{4.1e+20}, \num{1.0e+21}, \num{2.0e+21}, 
\num{3.1e+21}, \num{4.1e+21}, \num{5.1e+21}, \num{6.1e+21}, \num{7.2e+21}) $\mathrm{cm^{-2}}$. We show in grayscale the column density 
values below \num{1.5e+20} $\mathrm{cm^{-2}}$ to highlight the halo gas. 
 Right: MeerKAT H\,{\sc i} velocity field of NGC 4945. 
 The contour levels are 
$\mathrm{V_{sys} \pm 180~km~s^{-1}}$ in step of $\mathrm{30~km~s^{-1}}$, where 
$\mathrm{V_{sys} = 565~km~s^{-1}}$ and is indicated by the green contour. The red crosses indicate the kinematic center derived by TiRiFiC. 
The red circles at the bottom left corner of each panel show the size of the beam  (FWHM = \SI{60}{\arcsecond}).}
\label{fig:NGC4945-mom}
\end{figure*}

To get a map of the integrated \HI\ intensity and velocity field of NGC~4945, we calculate the zeroth- and the first-moment of its spectrum. 
To locate an area containing only genuine emission, we create a mask by smoothing the data cube to a resolution 
of \SI{75}{\arcsecond}. This cube is then used as a mask for the 
\SI{60}{\arcsecond} cube to select pixels above a set level of 3 times the rms noise of the lower resolution cube. The moment 
calculations are done with the masked data cube. We show the integrated intensity map (moment zero) and velocity field (moment one) of the H\,{\sc i} emission 
in Fig.~\ref{fig:NGC4945-mom}.  The moment zero map reveals previously undetected faint H\,{\sc i} emission, extending out 
to a major axis diameter of \SI{32}{\arcminute} and a minor axis diameter of \SI{18}{\arcminute} (62\,$\times$\,35\,kpc), 
measured at a position angle of 45$\degree$, and at a mass surface density level of 0.02$\mathrm{~M_{\odot}\,pc^{-2}}$. An
illustration of how these numbers are derived is shown in the Appendix. The distribution is 
clearly asymmetric, with the south-eastern side having much more faint extended emission than the rest of the galaxy. 
\\

The observations by the LVHIS survey \citep{2018MNRAS.478.1611K} provide the most sensitive 
H\,{\sc i} maps of NGC~4945 in the literature. 
On the receding side, the radial extent of the moment zero map of our MeerKAT H\,{\sc i} observation is similar to the extent found by LVHIS as 
it also ends at the edge of the optical disc as shown in Fig.~\ref{fig:NGC4945-mom}. 
However, on the approaching side, our map radially extends slightly further than that of LVHIS. In addition, we find significantly more gas in the lower 
left quadrant of the galaxy than what was detected by the LVHIS. This is due to our improved sensitivity, going down to a $3\sigma$ 
column density level of \num{5e+18} $\mathrm{cm^{-2}}$ at a 44 $\mathrm{km~s^{-1}}$ velocity resolution and a \SI{60}{\arcsecond}  beam, 
compared to \num{1.2e+19} $\mathrm{cm^{-2}}$ for LVHIS. 
The finger-like structures seen on both sides of the major axis of NGC~4945 were already apparent on the LVHIS map, 
though they are more clearly visible here. The area where H\,{\sc i} absorption occurs appear as blanked pixels 
in Fig.~\ref{fig:NGC4945-mom}. There is a central concentration of high-density gas extending 
symmetrically with respect to the minor axis towards each side of the galaxy. 
The disc velocity field shows the regular spider pattern, which indicates a rotationally-supported system. There are kinks in the iso-velocity contours, 
which have usually been attributed to non-circular motions caused by bars, spiral arms or star formation. 
In addition, irregular patterns are seen in the South-Eastern side where the faint extended emission is detected. Finally, the iso-velocity 
contours appear to be twisted on the approaching and the receding sides, indicating a warp-like morphology. 

\subsection*{\HI\ absorption} 
\begin{figure} 
\begin{tabular}{l}
\includegraphics[scale=0.6]{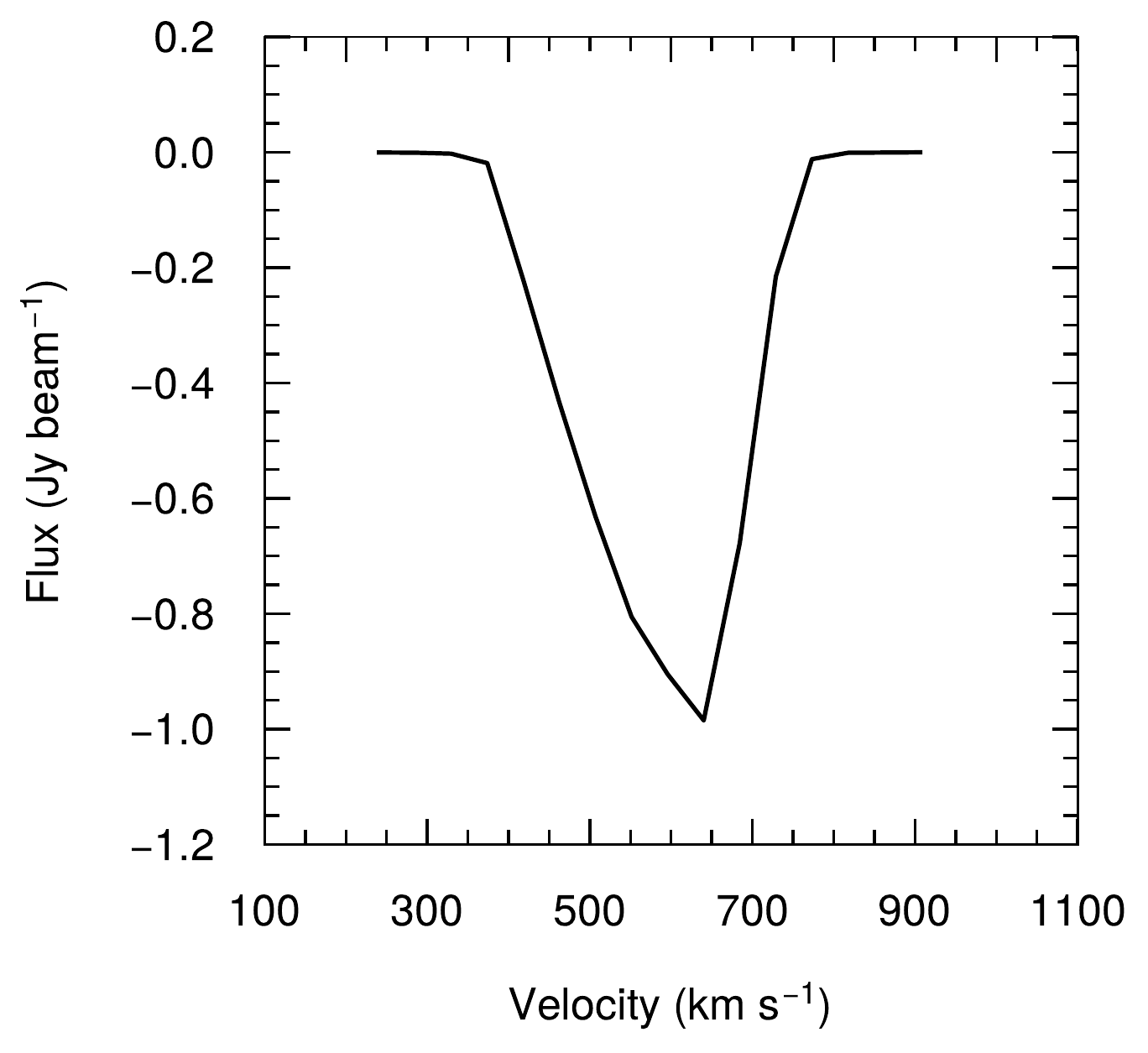} 
\end{tabular}
\caption{H\,{\sc i} absorption profile at the central peak of the nuclear continuum in NGC~4945.} 
\label{fig:NGC4945-abs}
\end{figure}

\begin{figure*} 
\begin{tabular}{l}
\includegraphics[scale=0.65]{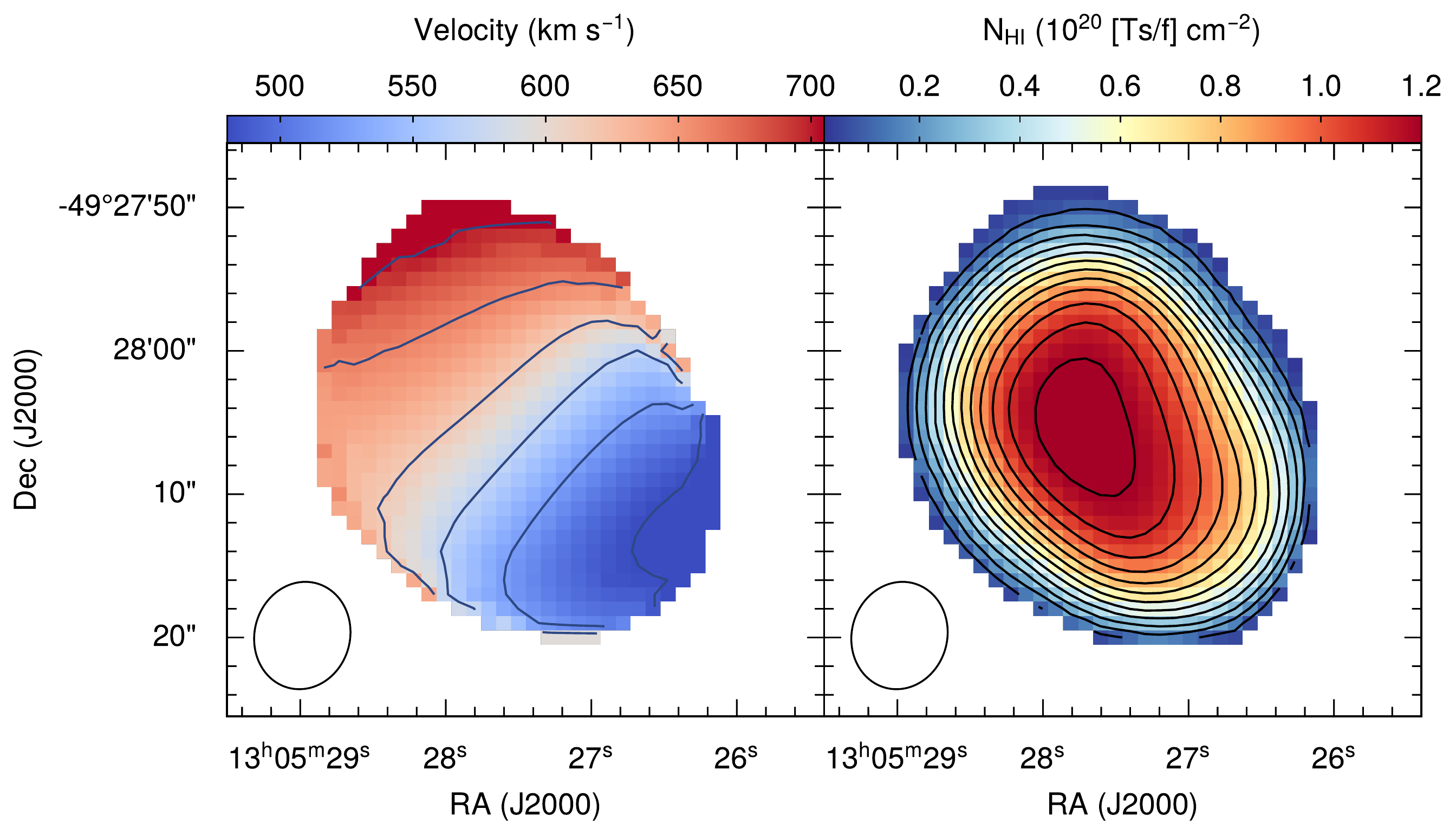}
\end{tabular}
\caption{Left: velocity field of the H\,{\sc i} absorption in the central region of NGC~4945, using the data cube at \SI{7.5}{\arcsecond} $\times$ \SI{6.4}{\arcsecond} 
resolution.  The contour levels are (480, 525, 570, 615, 660, 705)  $\mathrm{~km~s^{-1}}$. Right: column density map of the H\,{\sc i} absorption in the central region of NGC~4945, using the data cube at \SI{7.5}{\arcsecond} $\times$ \SI{6.4}{\arcsecond} 
resolution.  The contour levels are 0.1 to 1.2 in step of 0.1 in units of $\mathrm{~10^{20}~[Ts/f]~cm^{-2}}$.  The  black ellipses in each panel show the beam.} 
\label{fig:NGC4945-abs-velfield}
\end{figure*}

\HI\ absorption is seen against the bright radio continuum emission in the nuclear region of NGC~4945. 
Most notably, the \HI\ absorption is seen throughout the entire velocity range of the \HI\ emission of the galaxy. 
As shown in Fig.~\ref{fig:NGC4945-ch} \HI\ absorption is present even in channels without \HI\ emission. 
We show the \HI\ absorption line at the peak of the nuclear continuum emission in Fig.~\ref{fig:NGC4945-abs}. 
It has a peak velocity of 640 km\,s$^{-1}$, a peak flux-density of --1 Jy/beam, and 
a wide velocity extent ranging from 375\,km~s$^{-1}$ to 777\,km~s$^{-1}$. The profile has a $W_{50}$ of 
226 $\mathrm{km~s^{-1}}$ and $W_{20}$ of 315 $\mathrm{km~s^{-1}}$. This is within the range of broad \HI\ absorption profiles 
as categorised by \citet{2015A&A...575A..44G}. They found that broad \HI\ absorption profiles tend 
to be asymmetric and most likely arise from either a highly turbulent \HI\ gas or a fast rotating ring. 
Here also we find the \HI\ absorption profile of NGC~4945 
to be broad and asymmetric. To quantify the asymmetry of the profile, we use the following 
relation by \citet{2015A&A...575A..44G}
\begin{equation}
    a_{v} = max\left( \dfrac{v_{FW20 R} - v_{HI peak}}{v_{HI peak} - v_{FW20 B}}, 
    \left(\dfrac{v_{FW20 R} - v_{HI peak}}{v_{HI peak} - v_{FW20 B}}\right)^{-1} \right)
\end{equation}
, where $v_{HI peak}$ is the velocity at the position of the peak, and $v_{FW20R}$ $v_{FW20B}$ are the velocities at 20\% of the peak flux on the receding
and the approaching side of the profile with respect to its peak position, respectively. 
As explained by \citet{2015A&A...575A..44G}, the maximum value between the velocity ratio and its reciprocal ensures
that the asymmetry value is always greater than 1, independent of whether the line is skewed to the left or to the right. 
We find $a_{v}$ = 2.37. This value is consistent with the asymmetric parameter values found by \citet{2015A&A...575A..44G} for 
galaxies with FW20 $\geq$ 300 $\mathrm{km~s^{-1}}$. In Fig.~\ref{fig:NGC4945-abs-velfield}, we show the velocity field of the \HI\ absorption in 
the central region of NGC 4945 using the high resolution data cube at \SI{7.5}{\arcsecond} $\times$ \SI{6.4}{\arcsecond}. 
The velocity field indicates a clear solid-body rotation pattern, very similar to the radio recombination lines (RRL) velocity field 
derived by \citet{2010A&A...517A..82R} and the \HI\ absorption velocity field by \citet{2001A&A...372..463O}.
Thus, the broad \HI\ absorption lines in NGC 4945 cannot be caused by a highly turbulent gas but rather by a fast rotating ring of neutral gas 
around a central AGN. As explained by \cite{1993ApJ...402L..41K}, there would be no systematic shift in velocity with position if the broad lines were 
caused by gas moving at random velocity. However, outflows or inflows may be 
responsible for the asymmetry in the absorption lines of the galaxy. See also similar cases in NGC 1808 \citep{1993ApJ...402L..41K} and other 
nearby spiral galaxies \citep{1996ASPC..106..238K}. 
Another evidence for the existence of a fast rotating ring around the central AGN of NGC 4945 includes the detection of 
a megamaser with a broad velocity range (400 $\mathrm{km~s^{-1}}$ to 1100 $\mathrm{km~s^{-1}}$) by \citet{2016A&A...592L..13H}. 

We show the column density map of the H\,{\sc i} absorption, $N_{HI}$, in the central region of NGC~4945 in Fig.~\ref{fig:NGC4945-abs-velfield}. The column density 
was derived using the following equation
\begin{equation}
N_{H\, \textsc i} = \num{1.82243e18} ~ Ts ~ \dfrac{F_{H\, \textsc i,~abs}}{f F_{H\, \textsc i,~cont}} ~ \Delta v,
\end{equation}
where $Ts$ is the spin temperature of hydrogen atoms in K, $F_{H\, \textsc i,~abs}$  is the flux from the \HI\ absorption lines in $Jy~beam^{-1}$, 
$F_{H\, \textsc i,~cont}$ is the flux from the 21-cm radio continuum in $Jy~beam^{-1}$, $f$ is the covering factor, and $\Delta v$ is the channel width in $\mathrm{km~s^{-1}}$. 
We find $N_{H\, \textsc i}$  = \num{1.7e+18} - \num{1.3e+20} [Ts/f $\mathrm{cm^{-2}}$].
The spin temperature is a poorly constrained parameter and depends on a lot of excitation mechanisms such as the cosmic microwave background radiation (CMB), collisional excitation, and Ly$\alpha$ radiation \citep{1959ApJ...129..536F}. 
The column density of the  \HI\  emission in the central region of NGC 4945 is well above the threshold value for cold neutral medium formation (CNM), $N_{H\, \textsc i}$~$\gtrsim$~\num{2e+20}~$\mathrm{cm^{-2}}$ \citep{2011ApJ...737L..33K}. 
Thus we expect the \HI\ in the central region of NGC 4945 to be predominantly in the cold phase with Ts $\lesssim$ 500 K \citep{2011ApJ...737L..33K}. This is supported by the presence of molecular gas in NGC 4945 as reported by \cite{2001A&A...372..463O} and the high rate of nuclear star formation, indicating an enhanced atomic-to-molecular conversion process.

\subsection{Kinematic modelling}\label{sec:modelling} 
To model the H\,{\sc i} gas distribution and kinematics in NGC~4945, we use the 3D tilted ring fitting software 
FAT \citep{2015MNRAS.452.3139K} and TiRiFiC \citep{2007AA...468..903J}. FAT is built upon TiRiFiC but was designed to be run in a fully automated manner. 
FAT and TiRiFiC model a galaxy as a set of concentric rings with kinematic and orientation parameters that can 
either be varied or held fixed with radius. Each ring is characterised by its centre position, inclination angle with 
respect to the line of sight, position angle of the receding major axis, systemic velocity, rotation velocity, velocity dispersion, 
vertical thickness, and surface brightness distribution. The disc can be segmented and fitted independently, allowing the 
user to e.g., fit the approaching and the receding halves independently. TiRiFiC can be used to model non-axisymmetric features and 
large-scale motions such as those induced by the presence of bars. This can be achieved by including harmonic distortions in velocity and/or 
surface brightness distribution while modelling.

\subsubsection{Modelling strategy} 
To model NGC~4945, first, we make a base model with FAT. Then, we use the output model of FAT as input for TiRiFiC to refine the model 
and identify features not captured by FAT. We start with the simplest possible model, i.e., a model 
with the geometric parameters, systemic velocity, and velocity dispersion held fixed with radius, but allowing 
for the surface brightness and the rotation velocity to vary as a function of radius. We visually compare the resulting model 
with the data and add/vary one or more parameters if the simpler model does not adequately describe the data. As briefly mentioned before, 
we excluded the faint emission in the fitting as we failed to properly model it due to our coarse velocity 
resolution of 44 $\mathrm{km~s^{-1}}$. We will discuss more the properties of the anomalous gas in Section~\ref{sec:halogas}. 
Thus, in addition to applying a mask based on the \SI{75}{\arcsecond} data as 
described in the previous section, we create an elliptical region surrounding the main disc of the galaxy 
and use the MIRIAD task IMMMASK to mask out areas outside the ellipse. We put zeros in the areas that have been masked out. The masked data cube 
is then fed into TiRiFiC for the modelling. Note that due to the presence of the central absorption, which has been masked out, 
the fitting in the central part (about an arcminute in radius) is uncertain. 
Thus we refrain from doing mass modelling.

\subsubsection{Model parameters}
\begin{figure}
 \includegraphics[scale=0.31]{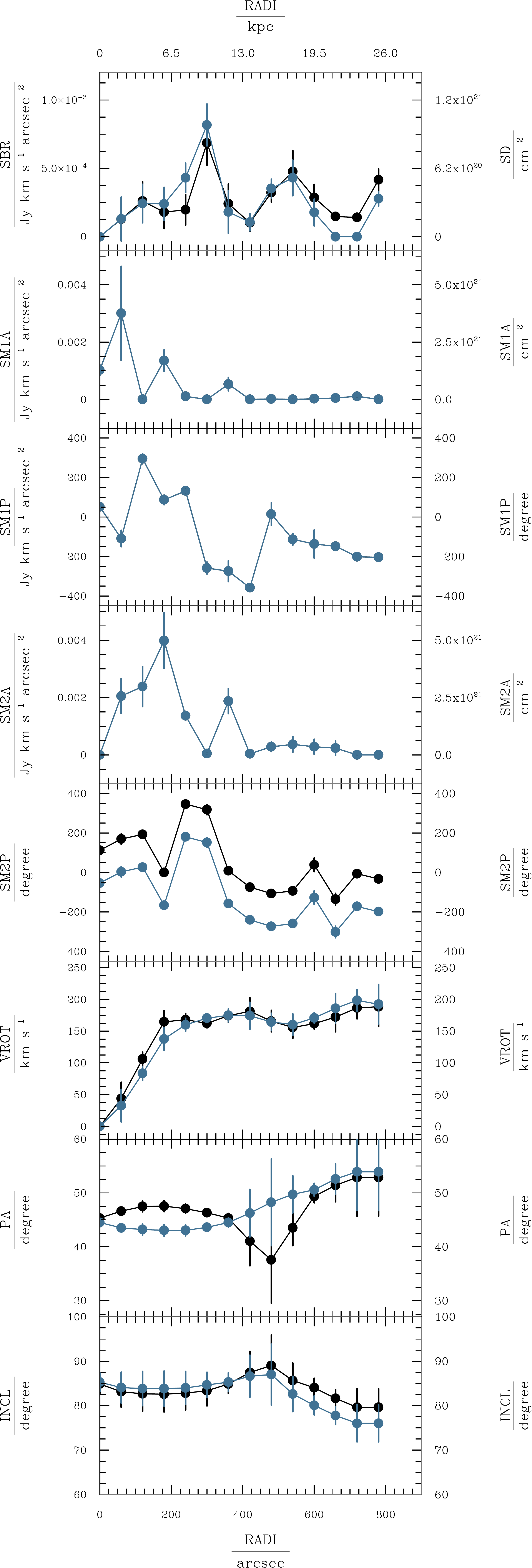}
 \caption{{\sc TiRiFiC} model parameters of NGC~4945 that vary with radius, they are tabulated in Table~\ref{tab:title}. 
 Black: approaching side, blue: receding side. For SM1A, SM1P, and SM2A, 
 the black and the blue curves overlap.}
\label{fig:gal_modelpars}
\end{figure}

Our final best-fitting model is a model where we fit the approaching and receding sides separately as described below. 
\begin{enumerate}[label*=\textbf{$\bullet$},itemsep=0.5em]
\item The following parameters have been kept constant with radius:
\begin{enumerate}[topsep=12pt]
   \item centre position: XPOS, YPOS, XPOS\_2, YPOS\_2
   \item systemic velocity: VSYS, VSYS\_2
   \item global dispersion: CONDISP (this does not vary with radius by default, and should not be confused with the 
   SDIS parameter which can be varied with radius)
   \item disc thickness: Z0, Z0\_2
\end{enumerate}
\item The following parameters have been allowed to vary with radius:
\begin{enumerate}[topsep=12pt]
    \item rotation velocity: VROT, VROT\_2
    \item surface brightness: SBR, SBR\_2
    \item amplitude of harmonic distortions in surface brightness (first and second order): SM1A, SM1A\_2; SM2A, SM2A\_2
    \item phase of harmonic distortions in surface brightness (first and second order): SM1P, SM1P\_2; SM2P, SM2P\_2
    \item inclination: INCL, INCL\_2
    \item position angle: PA, PA\_2
\end{enumerate}
\end{enumerate}
We show the parameters that are kept constant with radius in Table~\ref{tab:model_prop}. The ones that vary with radius are shown in 
Fig.~\ref{fig:gal_modelpars} and Table~\ref{tab:title}. 
\begin{table}
    \centering
    \caption{Tilted-ring parameters not varying with radius.}
 \begin{small}
    \begin{tabular*}{0.45\textwidth}{l @{\extracolsep{\fill}}lrl}
       \multicolumn{4}{p{0.45\textwidth}}{NGC~4945 TiRiFiC model parameters that are constant with radius}\\
       \hline
       \hline
	Parameter & Symbol & \multicolumn{1}{c}{Value} & Unit\\
       \hline
        \multicolumn{1}{p{0.09\textwidth}}{\multirow{2}{*}{\parbox{1.5cm}{Model centre (J2000)}}} & XPOS & $13.0^\mathrm{h}\, 05.0^\mathrm{ m}\,28.0^\mathrm{s} \, \pm\, 00.0^\mathrm{s}$&\\
	& YPOS & $-49.0^\mathrm{d}\,28.0^\mathrm{m}\, 05.8^\mathrm{s}\, \pm \, 00.4^\mathrm{s}$ &  \\
	\multicolumn{1}{p{0.09\textwidth}}{\multirow{1}{*}{\parbox{1.5cm}{Systemic Velocity}}} & VSYS & $565\,\pm\,6$ & $\mathrm{km}\,\mathrm{s}^{-1}$ \\
	&&&\\
	Thickness & Z0 & $42.6\arcsec\,\pm\,3.0\arcsec$ & \\
	Dispersion & CONDISP & $23.0\,\pm\,1.5$ & $\mathrm{km}\,\mathrm{s}^{-1}$\\
    \hline
    \end{tabular*}
    \end{small}
     
    \label{tab:model_prop}
 \end{table}
\textbf{General properties}:
\begin{enumerate}[label=\textbf{$\bullet$}]
\item We derive a flat rotation curve, typical of spiral galaxies. A fit to the following 
function \citep{2003MNRAS.346.1215B, 2008AJ....136.2782L} 
\begin{equation} 
 v_{rot}(r) = v_{flat}\left[1 - \exp \left(1-\dfrac{-r}{l_{flat}}\right) \right],
\end{equation}
gives the numbers quoted in Table~\ref{tab:vflat}, where 
$v_{rot}(r)$ is the rotation curve, $v_{flat}$ is the rotation velocity at the flat part of the rotation curve, 
$l_{flat}$ is the length scale over which $v_{rot}$ approaches $v_{flat}$ \citep{2008AJ....136.2782L}.  
\item The position angle and inclination clearly indicate an outer warp in the approaching and the receding sides. 

\item The surface brightness profile decreases toward the centre due to the presence of the absorption. 
It appears to be disturbed (irregular), most likely due to a bar. 
\end{enumerate}

\textbf{Properties of the kinematic model}: The approaching side shows a steeper velocity gradient than the receding side. This is already apparent in the velocity field. 
However, their maximum rotation velocities agree within the uncertainties. In addition, both sides have an outer warp. Moreover, there are differences in the radial variations 
of the surface brightness distribution. The receding side appears to be brighter in the inner discs and fainter in the outer discs compared to the approaching side. 
The amplitude and the phase of the first-order harmonic distortions in surface brightness for the two sides are similar, though.  

\begin{table*}
\caption {TIRIFIC model parameters of the NGC~4945 \HI\ disc that vary with radius, also shown in Fig.~\ref{fig:gal_modelpars}.} \label{tab:title} 
\setlength{\tabcolsep}{1pt} 
\renewcommand{\arraystretch}{1.2}
\resizebox{\textwidth}{!}{
\begin{adjustbox}{angle=0}
\begin{tabular}{ccccccccccccccccc}
\hline \hline
RADI & SBR & ERR\_SBR & SM1A & ERR\_SM1A & SM1P & ERR\_SM1P & SM2A & ERR\_SM2A & SM2P & ERR\_SM2P & VROT & ERR\_VROT & PA & ERR\_PA & INCL & ERR\_INCL \\
\hline
0. & 4.8e-08 & 4.9e-09 & 1.0e-03 & 1.8e-04 & 51.6 & 6.9 & 1.9e-07 & 3.5e-08 & 1.1e+02 & 8.5 & 0. & 0. & 45.3 & 0.8 & 84.9 & 2.1 \\
60. & 1.3e-04 & 1.6e-04 & 3.0e-03 & 1.6e-03 & -109. & 41.4 & 2.1e-03 & 6.0e-04 & 1.7e+02 & 26.2 & 43.8 & 25.7 & 46.6 & 0.8 & 83.1 & 3.5 \\
120. & 2.6e-04 & 1.4e-04 & 4.5e-06 & 5.1e-06 & 295. & 25.2 & 2.4e-03 & 6.9e-04 & 1.9e+02 & 15.2 & 106.2 & 10.7 & 47.5 & 1. & 82.7 & 3.9 \\
180. & 1.8e-04 & 1.2e-04 & 1.4e-03 & 3.7e-04 & 86.5 & 24. & 4.0e-03 & 9.7e-04 & 7.9e-02 & 13.7 & 164.6 & 17.9 & 47.6 & 1. & 82.6 & 4. \\
240. & 2.0e-04 & 1.1e-04 & 1.1e-04 & 9.5e-05 & 132. & 8.3 & 1.4e-03 & 1.5e-04 & 3.5e+02 & 13.2 & 168.1 & 9.7 & 47.1 & 0.8 & 82.8 & 3.8 \\
300. & 6.9e-04 & 1.6e-04 & 5.1e-07 & 4.8e-07 & -258.5 & 30.2 & 5.0e-05 & 1.0e-05 & 3.2e+02 & 25.6 & 162.3 & 7.1 & 46.3 & 0.5 & 83.4 & 3.4 \\
360. & 2.4e-04 & 1.4e-04 & 5.3e-04 & 2.3e-04 & -274.4 & 52.1 & 1.9e-03 & 4.3e-04 & 8.5e+00 & 5.4 & 174.3 & 10.3 & 45.3 & 0.8 & 84.9 & 2.1 \\
420. & 1.0e-04 & 6.4e-05 & 1.1e-06 & 1.5e-06 & -358.2 & 7. & 4.0e-05 & 4.7e-05 & -7.4e+01 & 16.4 & 181.3 & 21.3 & 41. & 4.6 & 87.5 & 4.7 \\
480. & 3.2e-04 & 6.9e-05 & 1.9e-05 & 3.5e-05 & 14.3 & 57.1 & 2.8e-04 & 1.6e-04 & -1.1e+02 & 16.5 & 165.9 & 16.9 & 37.6 & 8. & 89. & 6.8 \\
540. & 4.8e-04 & 1.5e-04 & 1.0e-06 & 1.1e-06 & -112.9 & 29.4 & 3.7e-04 & 2.7e-04 & -9.3e+01 & 12.2 & 155.8 & 17.1 & 43.5 & 3.3 & 85.7 & 3.9 \\
600. & 2.9e-04 & 9.6e-05 & 2.4e-05 & 2.4e-05 & -136.8 & 70.9 & 2.9e-04 & 2.6e-04 & 3.9e+01 & 34.4 & 161.8 & 8.6 & 49.4 & 1.2 & 84. & 2.1 \\
660. & 1.5e-04 & 2.3e-05 & 5.1e-05 & 2.9e-05 & -148.2 & 21.6 & 2.4e-04 & 2.4e-04 & -1.4e+02 & 29.3 & 172.2 & 23.2 & 51.4 & 3.1 & 81.6 & 2. \\
720. & 1.4e-04 & 1.8e-05 & 1.1e-04 & 2.1e-05 & -201.6 & 17.1 & 6.0e-07 & 8.5e-08 & -6.2e+00 & 11.7 & 186.7 & 17.2 & 52.9 & 7.2 & 79.6 & 4.2 \\
780. & 4.2e-04 & 8.0e-05 & 6.3e-07 & 1.1e-07 & -204. & 10.4 & 5.9e-07 & 9.6e-08 & -3.3e+01 & 8.2 & 188.5 & 31.1 & 52.9 & 7.2 & 79.6 & 4.2 \\
&&&&&&&&&&&&&&&&\\
&&&&&&&&&&&&&&&&\\
\hline \hline
RADI & SBR\_2 & ERR\_SBR\_2 & SM1A\_2 & ERR\_SM1A\_2 & SM1P\_2 & ERR\_SM1P\_2 & SM2A\_2 & ERR\_SM2A\_2 & SM2P\_2 & ERR\_SM2P\_2 & VROT\_2 & ERR\_VROT\_2 & PA\_2 & ERR\_PA\_2 & INCL\_2 & ERR\_INCL\_2 \\
\hline
0. & 4.9e-08 & 4.9e-09 & 1.0e-03 & 1.8e-04 & 51.6 & 6.9 & 1.9e-07 & 3.5e-08 & -53.7 & 8.5 & 0. & 0. & 44.5 & 0.8 & 85.3 & 2.1 \\
60. & 1.3e-04 & 1.6e-04 & 3.0e-03 & 1.6e-03 & -109. & 41.4 & 2.1e-03 & 6.0e-04 & 3.1 & 26.2 & 32.7 & 25.7 & 43.5 & 0.8 & 84.1 & 3.5 \\
120. & 2.4e-04 & 1.4e-04 & 4.5e-06 & 5.1e-06 & 295. & 25.2 & 2.4e-03 & 6.9e-04 & 26.9 & 15.2 & 83.5 & 10.9 & 43.2 & 1. & 83.9 & 3.9 \\
180. & 2.4e-04 & 1.2e-04 & 1.4e-03 & 3.7e-04 & 86.5 & 24. & 4.0e-03 & 9.7e-04 & -165.9 & 13.7 & 137.7 & 17.9 & 43.1 & 1.1 & 83.9 & 3.9 \\
240. & 4.3e-04 & 1.1e-04 & 1.1e-04 & 9.5e-05 & 132. & 8.3 & 1.4e-03 & 1.5e-04 & 180.3 & 13.2 & 160. & 10.1 & 43. & 1. & 84. & 3.7 \\
300. & 8.2e-04 & 1.5e-04 & 5.1e-07 & 4.8e-07 & -258.5 & 30.2 & 5.0e-05 & 1.0e-05 & 151.9 & 25.6 & 170.4 & 7.1 & 43.6 & 0.5 & 84.7 & 2.9 \\
360. & 1.8e-04 & 1.6e-04 & 5.3e-04 & 2.3e-04 & -274.4 & 52.1 & 1.9e-03 & 4.3e-04 & -157.5 & 5.4 & 174.9 & 10.2 & 44.5 & 0.8 & 85.3 & 2.1 \\
420. & 1.1e-04 & 6.3e-05 & 1.1e-06 & 1.5e-06 & -358.2 & 7. & 4.0e-05 & 4.7e-05 & -240.3 & 16.4 & 174.2 & 21.3 & 46.3 & 4.4 & 86.6 & 4.7 \\
480. & 3.5e-04 & 6.9e-05 & 1.9e-05 & 3.5e-05 & 14.3 & 57.1 & 2.8e-04 & 1.6e-04 & -272.7 & 16.5 & 164.5 & 13.7 & 48.3 & 8. & 87. & 6.8 \\
540. & 4.3e-04 & 1.3e-04 & 1.0e-06 & 1.1e-06 & -112.9 & 29.4 & 3.7e-04 & 2.7e-04 & -259.1 & 12.2 & 160.3 & 17.1 & 49.7 & 3.5 & 82.6 & 4. \\
600. & 1.8e-04 & 9.8e-05 & 2.4e-05 & 2.4e-05 & -136.8 & 70.9 & 2.9e-04 & 2.6e-04 & -127.2 & 34.4 & 170.9 & 8.3 & 50.6 & 1.2 & 80.1 & 2.1 \\
660. & 2.3e-07 & 1.4e-07 & 5.1e-05 & 2.9e-05 & -148.2 & 21.6 & 2.4e-04 & 2.4e-04 & -301.2 & 29.3 & 186.1 & 23.2 & 52.6 & 2.8 & 77.8 & 2. \\
720. & 4.7e-07 & 5.6e-08 & 1.1e-04 & 2.1e-05 & -201.6 & 17.1 & 6.0e-07 & 8.5e-08 & -172.2 & 11.7 & 198.5 & 17.1 & 53.9 & 7.2 & 76. & 4.2 \\
780. & 2.8e-04 & 5.3e-05 & 6.3e-07 & 1.1e-07 & -204. & 10.4 & 5.9e-07 & 9.6e-08 & -198.6 & 8.2 & 192.4 & 31.1 & 53.9 & 7.2 & 76. & 4.2 \\
\hline
\end{tabular}
\end{adjustbox}}
\end{table*}
\begin{table}
\centering
\caption{Flat part of the TiRiFiC rotation curve}
\begin{tabular}[t]{lcc}
\cline{2-3}
&$v_{flat}$&$l_{flat}$\\
&$\mathrm{km~s^{-1}}$ & kpc\\
\hline
Approaching side&177.5 $\pm$ 5.9&3.8 $\pm$ 0.6 \\
Receding side&188.0 $\pm$ 7.1&5.3 $\pm$ 0.8 \\
\hline
\end{tabular}
    \begin{tablenotes}
      \small
      \item $v_{flat}$: the rotation velocity at the flat part of the rotation curve, 
$l_{flat}$ is the length scale over which $v_{rot}$ approaches $v_{flat}$ \citep{2008AJ....136.2782L}. 
    \end{tablenotes}
 \label{tab:vflat}
\end{table}%
\subsubsection{TiRiFiC model vs data}
\begin{figure*}
 \includegraphics[scale=0.79]{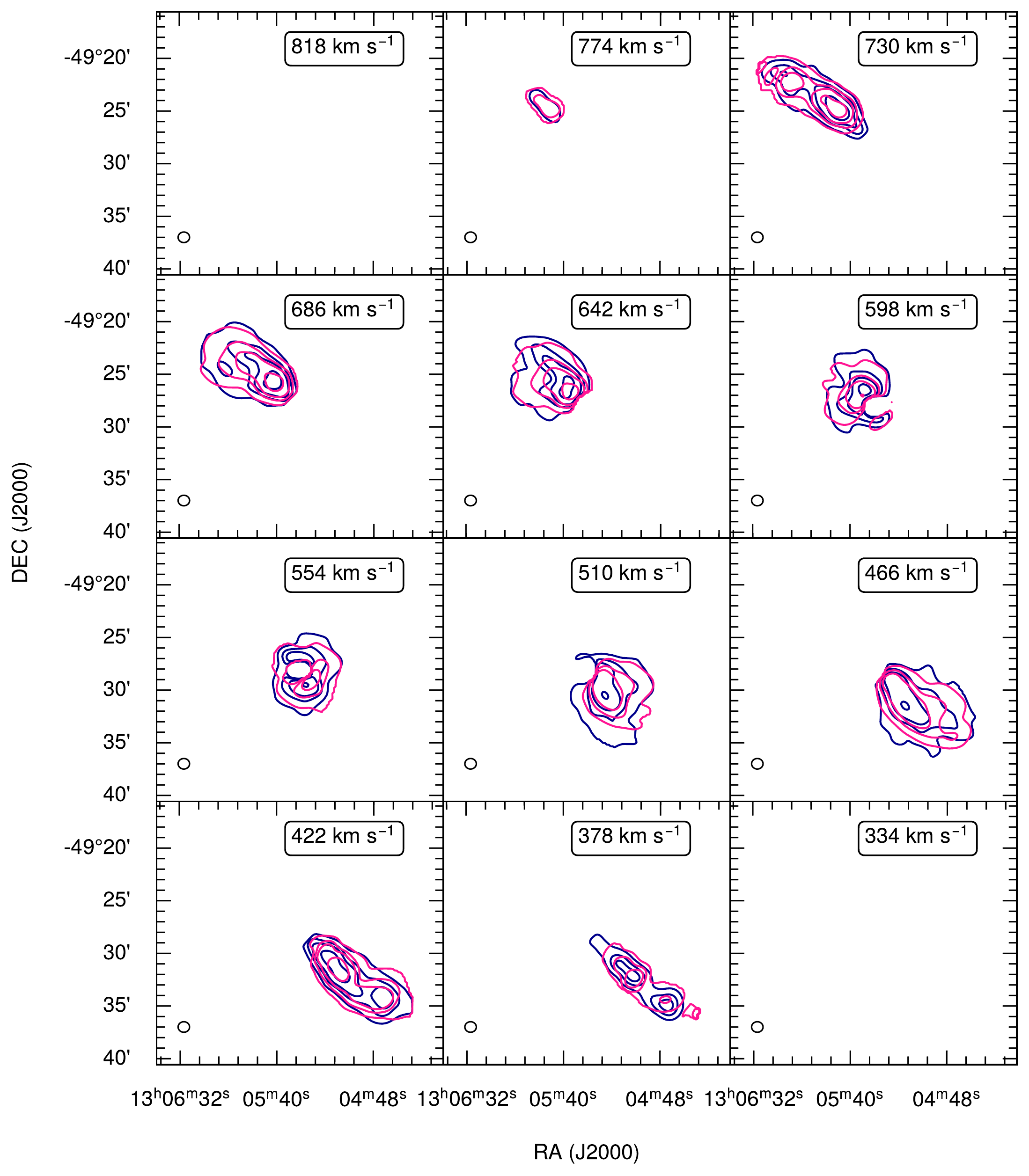}
 \caption{Comparison between the observed and the model \HI\ data cubes of NGC~4945. Here, only the gas from the main disc of NGC 4945 is shown since we could not fit 
 the gas in the halo as mentioned in the text. Blue: \HI\ channel maps of observed data cube; deep pink: channel maps of 
 TiRiFiC model data cubes. The contour levels are 0.0072, 0.0240, 0.0480, and 0.1200 $\mathrm{Jy~beam^{-1}}$. The black circles at the bottom left corner of 
 each panel show the size of the beam  (FWHM = \SI{60}{\arcsecond}). }
 \label{fig:modvsobschan}
\end{figure*}

\begin{figure*}
\begin{tabular}{cc}
\includegraphics[scale=0.64]{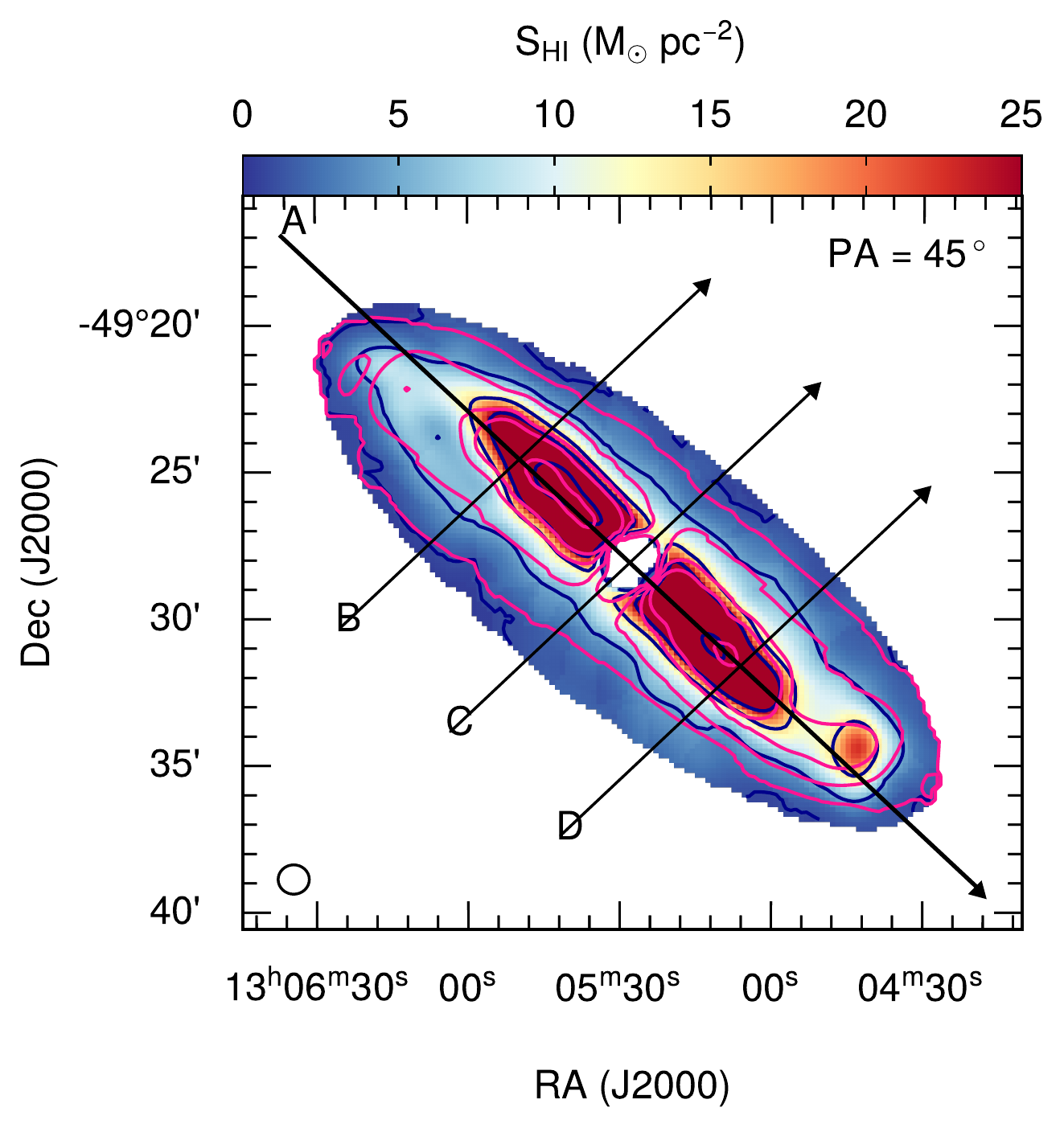}&
 \hspace*{-0.3cm}
\includegraphics[scale=0.64]{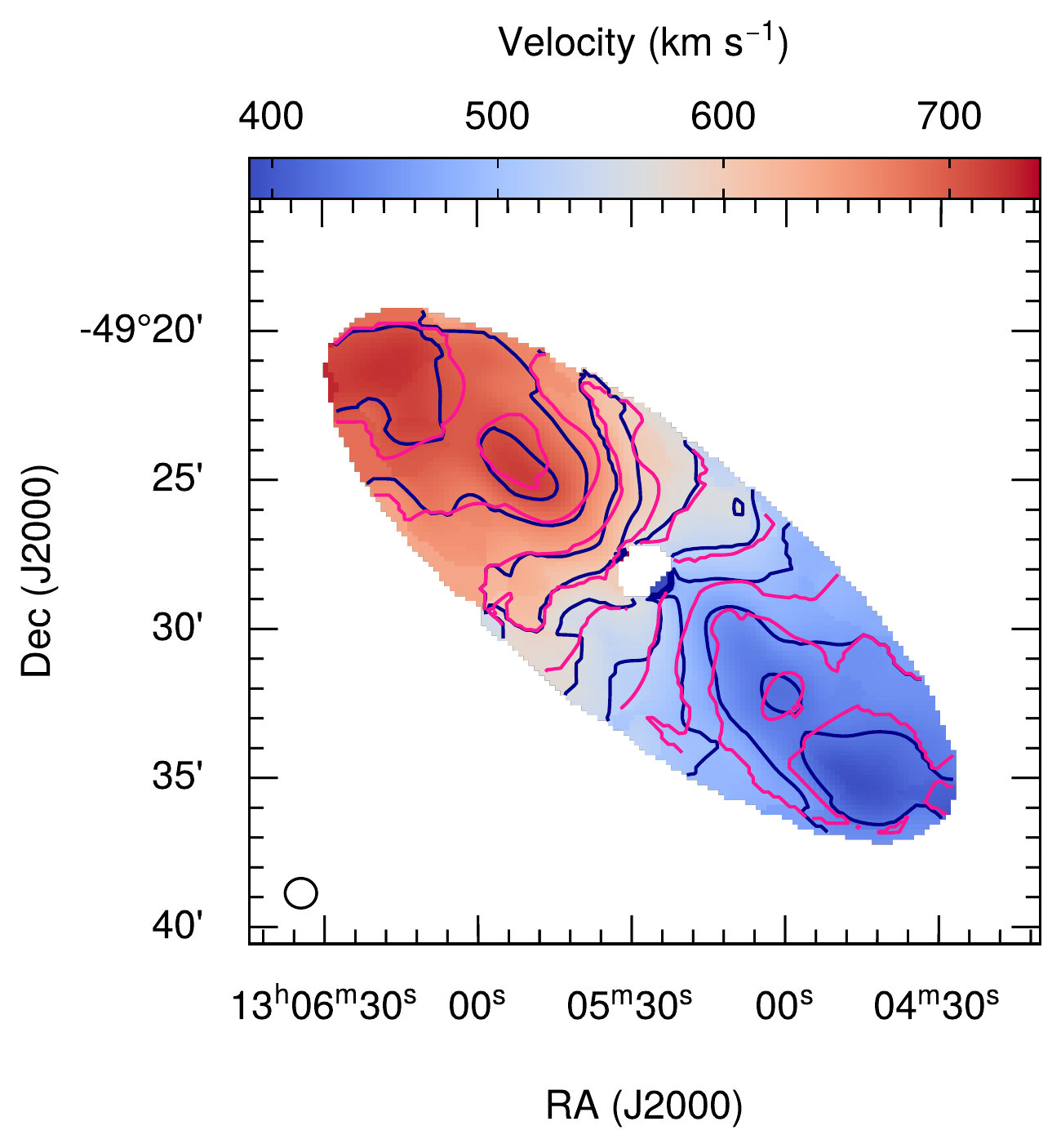}\\
 \includegraphics[scale=0.64]{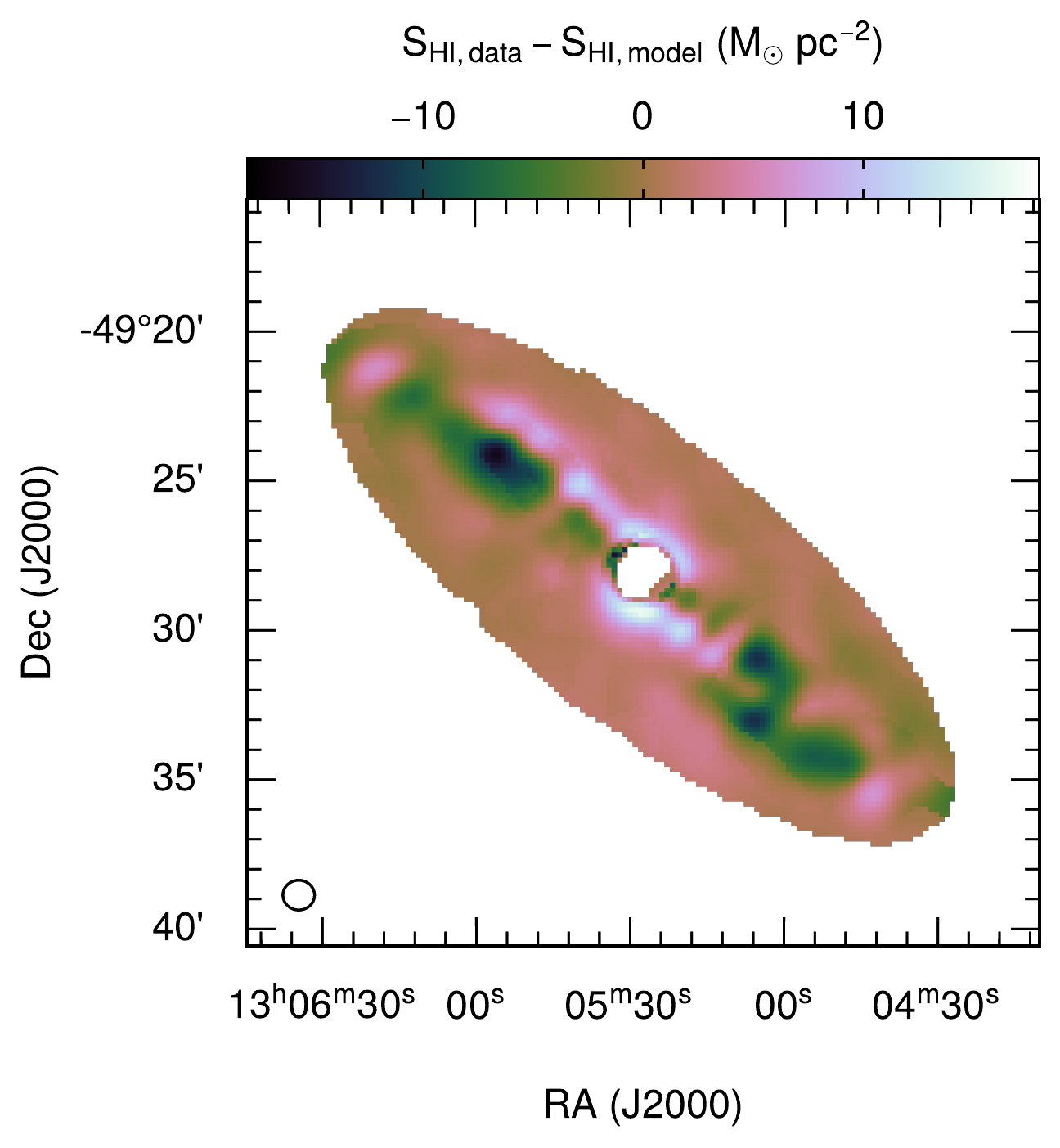} &
  \hspace*{-0.3cm}
 \includegraphics[scale=0.64]{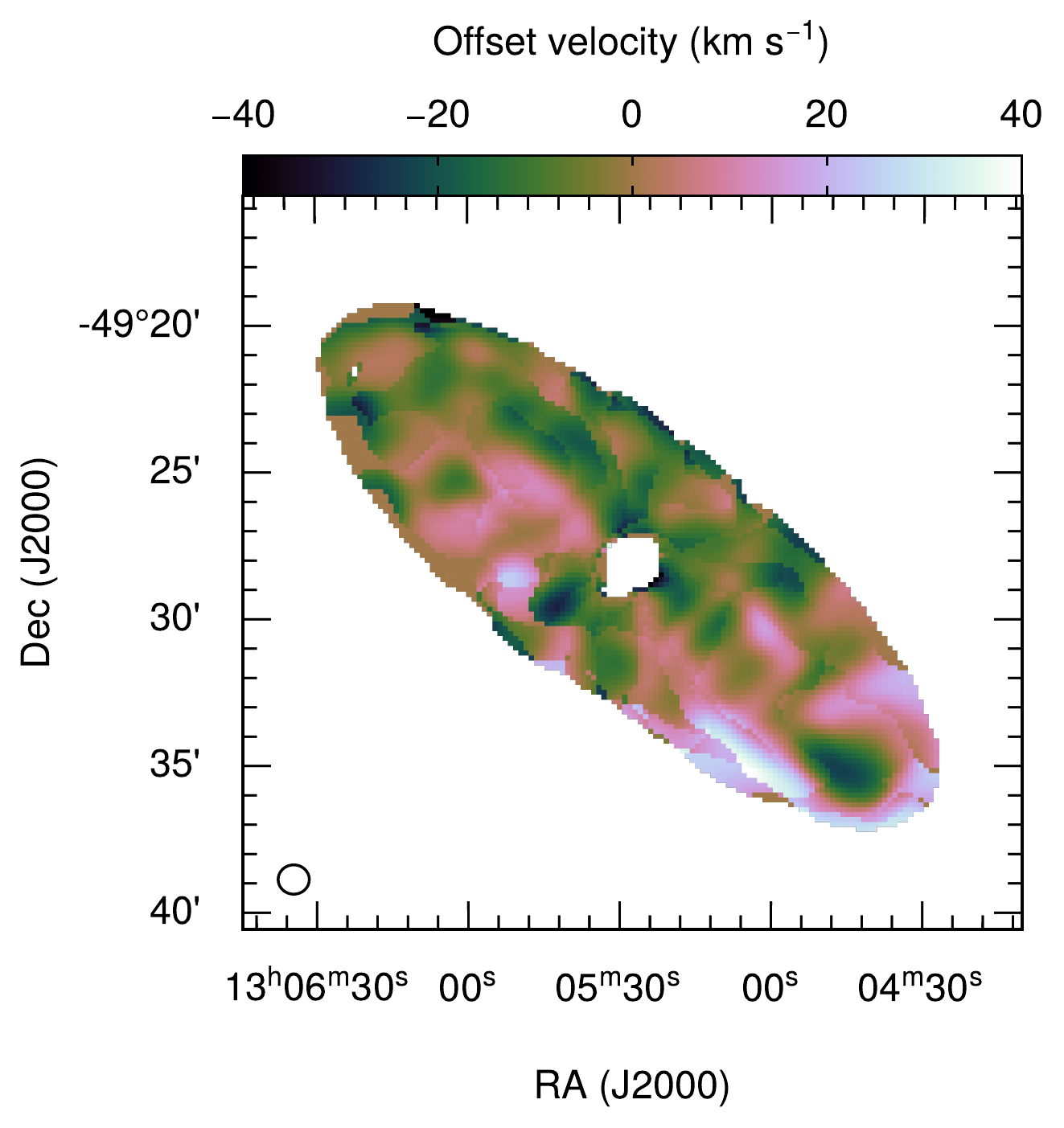}
 \end{tabular}
 \caption{Comparison between the NGC~4945 \HI\ moment maps from the observed and the model data cubes.  
 Only the gas from the main disc of the galaxy is shown as we failed to model the gas in the halo of NGC 4945.
 Top left: comparison between the observed \HI\ surface density map and the TiRiFiC model surface density map. Arrows A, B, C, and D show slices from which the position-velocity diagrams shown in Fig.~\ref{fig:modvsobspvd} and Fig.~\ref{fig:modvsobspvd1} are taken. The intersection between arrow A and arrow C represents the kinematic centre derived by TiRiFiC. The blue contours show the observed \HI\ surface density. The deep pink contours represent the surface density map from the TiRiFiC model data cube. The contour levels are 1, 5, 15, 25, and 55 $\mathrm{M_{\odot}~pc^{-1}}$. Top right: first moment maps;  the blue contours represent the observed velocity field, the deep pink contours show the model velocity field. 
 The contour 
 levels are $V_{sys}$ $\pm$ 175 $\mathrm{km}\,\mathrm{s}^{-1}$ in steps of 35 $\mathrm{km}\,\mathrm{s}^{-1}$ where $V_{sys}$ = 565 $\mathrm{km}\,\mathrm{s}^{-1}$. 
 The residuals are shown in the bottom panels of the Figure. Bottom left: the difference between the observed \HI\ surface density and the model \HI\ surface density. Bottom right: the difference between the observed 
 velocity field and the model velocity field. The circles at the lower left corners of each plot represent the size of the beam (FWHM = \SI{60}{\arcsecond}).}
 \label{fig:modvsobsmom}
\end{figure*}

\begin{figure*}
\begin{tabular}{cc}
 \includegraphics[scale=0.56]{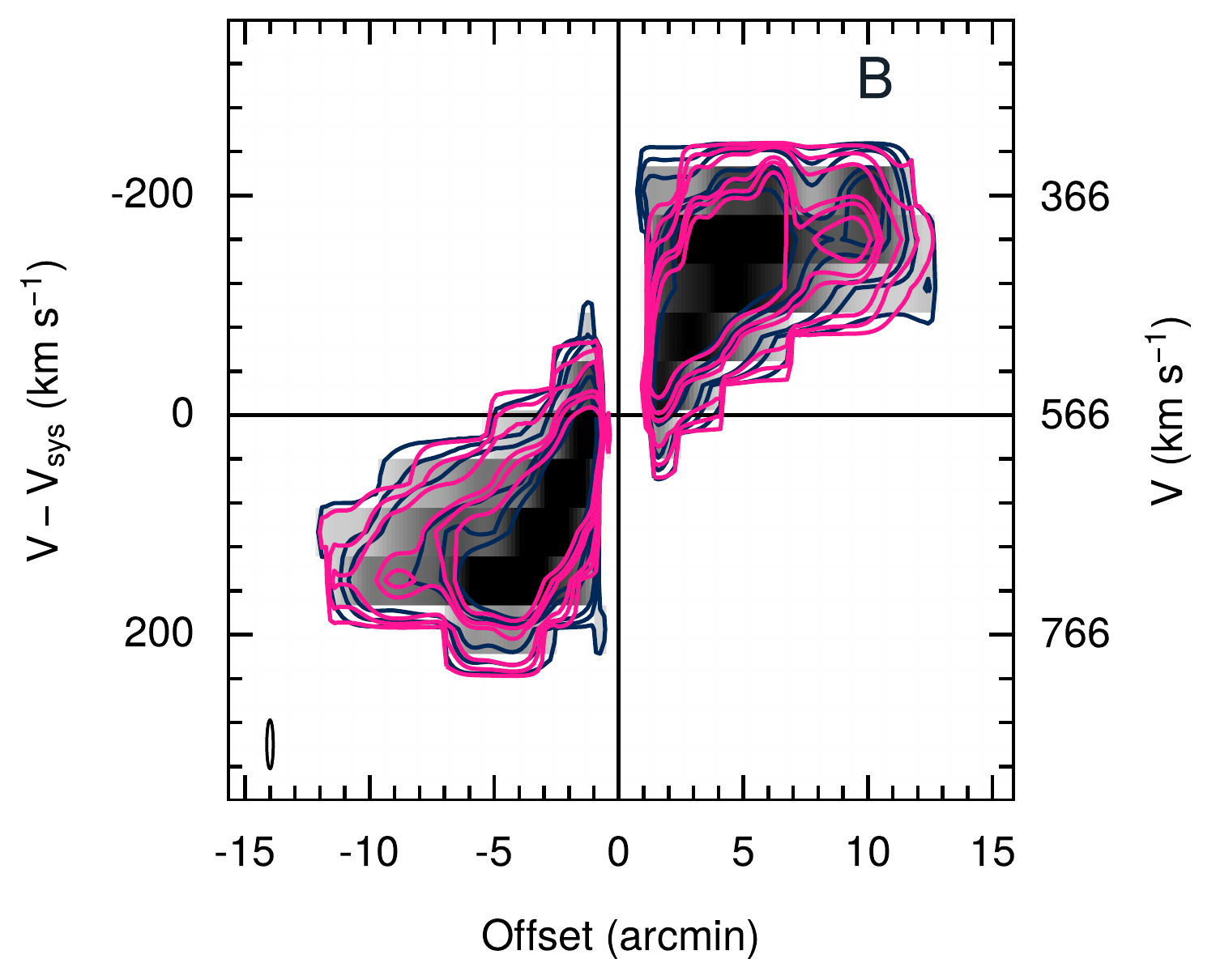}&
 \includegraphics[scale=0.56]{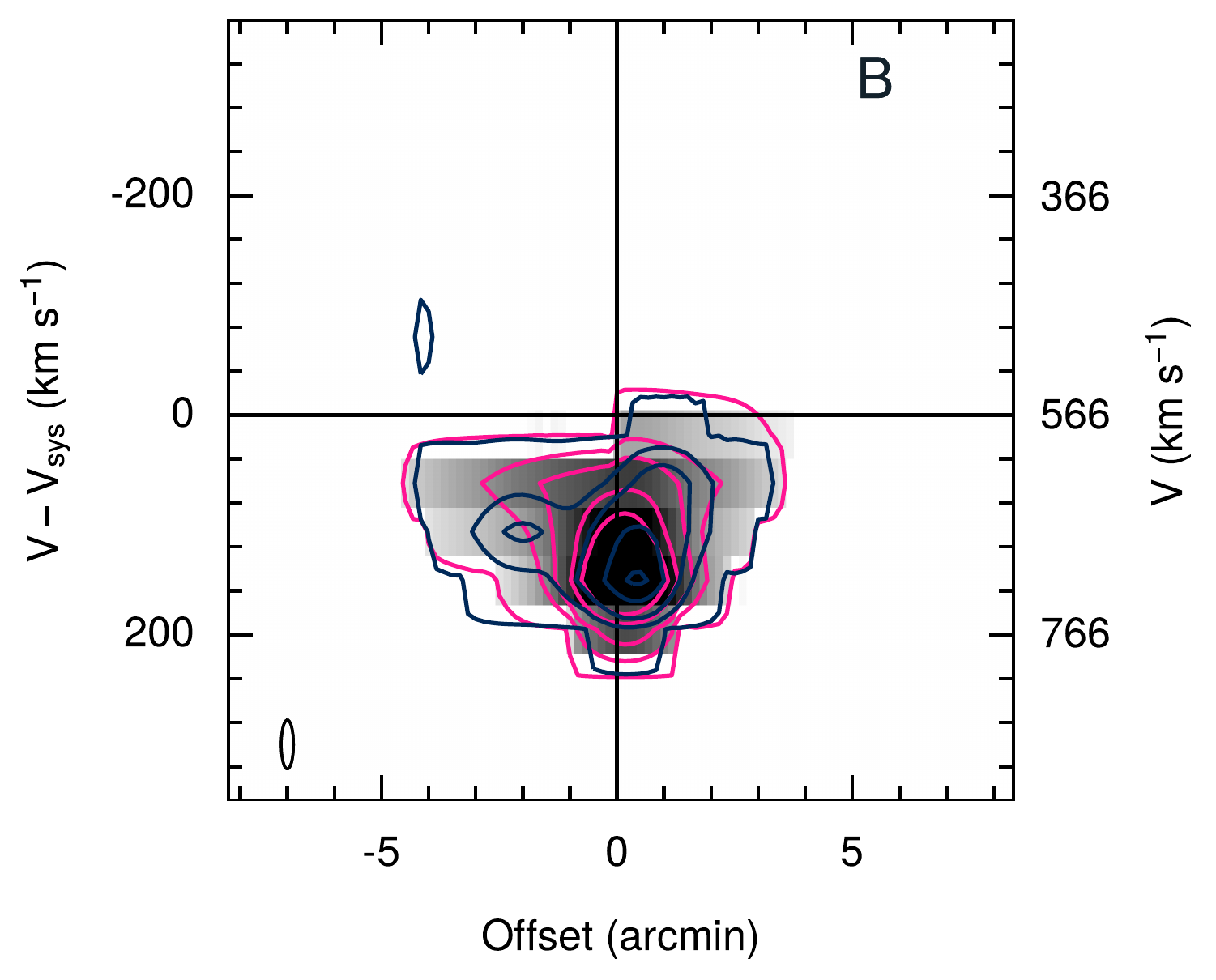}\\
 \includegraphics[scale=0.56]{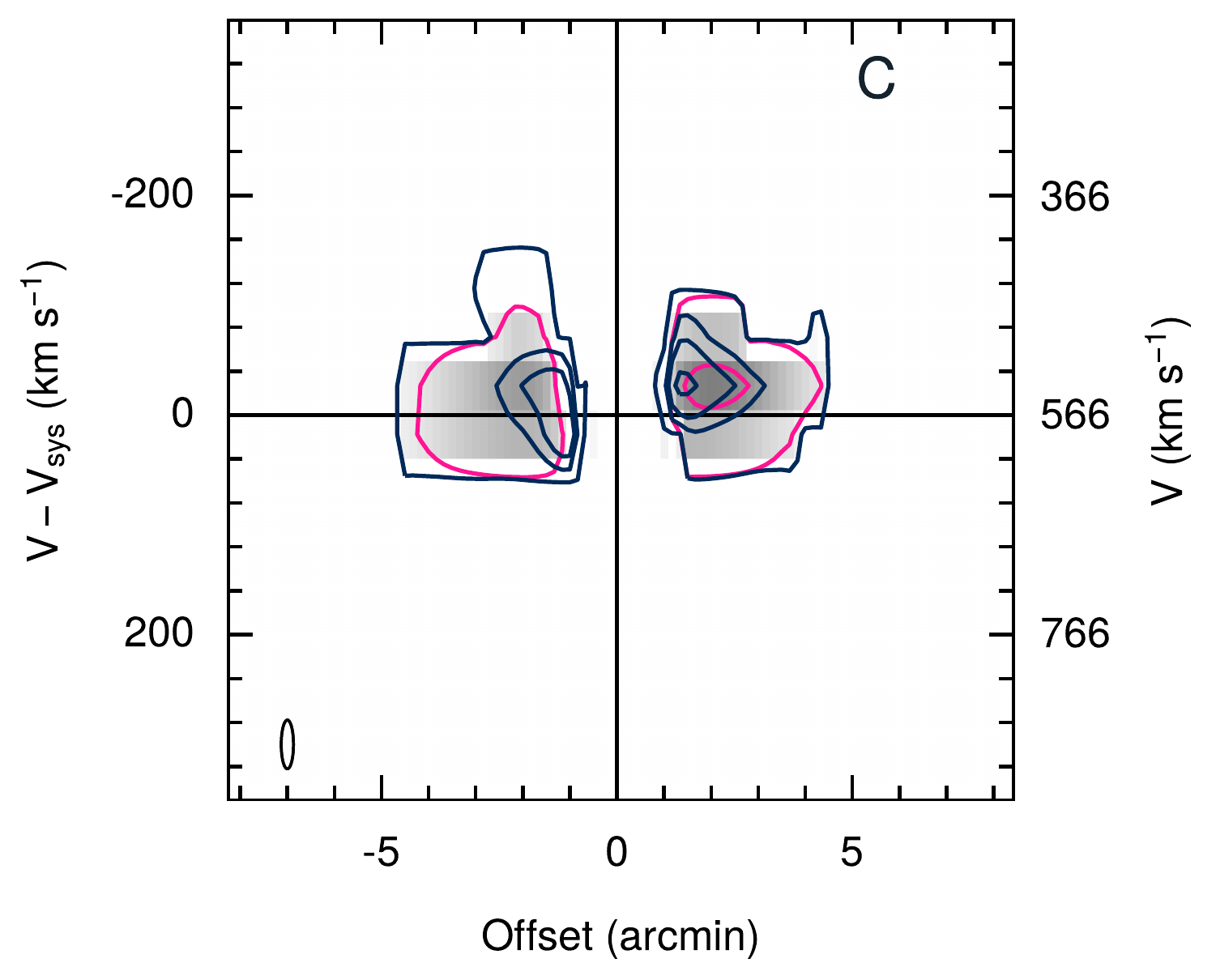}&
 \includegraphics[scale=0.56]{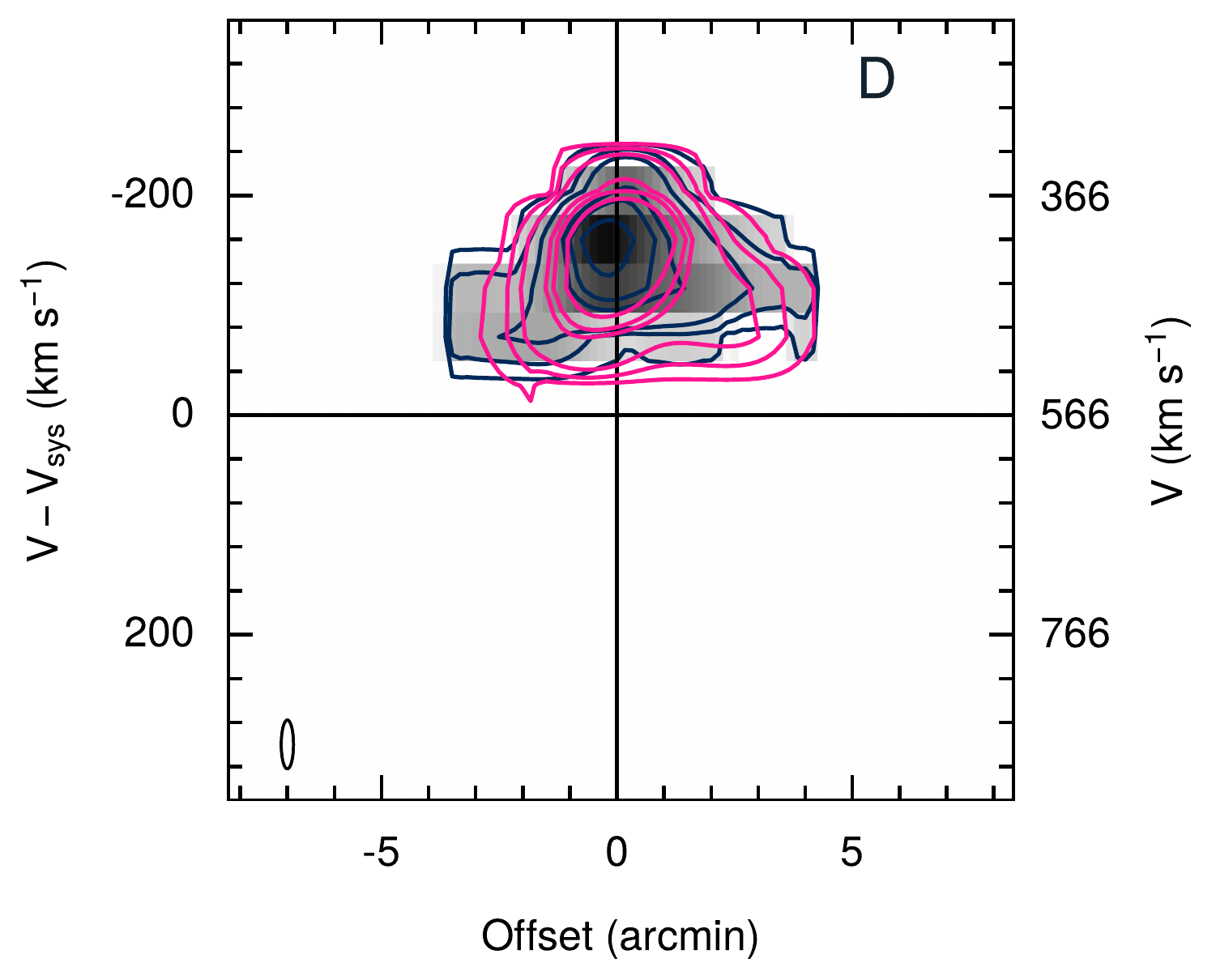} 
 \end{tabular}
 \caption{MeerKAT \HI\ position-velocity diagrams from the model and the observed data cubes of NGC~4945 taken 
 along slices A, B, C, and D as shown in Fig.~\ref{fig:modvsobsmom}.   Here also we only show the gas in the main disc of the galaxy since we could not model 
 the gas in the halo as mentioned in the text.
 Blue contours: observed data. Pink contours: the TiRiFiC model. The contour levels are (0.00072, 0.01200, 0.02400, 0.04800, 0.07200) $\mathrm{Jy~beam^{-1}}$. 
 The elongated ellipses at the lower left corners of the plots shows the size of the beam (FWHM = \SI{60}{\arcsecond}, along the x-axis)
and the channel spacing (along the y-axis).}
 \label{fig:modvsobspvd}
\end{figure*}

We compare the \HI\ channel maps from the TiRiFiC model data cube with the \HI\ channel maps from the observed data cube in Fig.~\ref{fig:modvsobschan}. 
Overall, the agreement is good, except close to the systemic velocity. This is expected due to the absorption in the centre and the effects of radial/streaming motions possibly
related to a bar, which are not included in the modelling. We compare the \HI\ moment maps from the TiRiFiC model with the \HI\ moment maps from the observations 
in Fig.~\ref{fig:modvsobsmom}.  The model recovers the overall 
gas distributions and kinematics. Local deviations exist, though. We also show the residual \HI\ moment maps obtained by subtracting 
the model from the data in Fig.~\ref{fig:modvsobsmom}. 
The residual velocity field has a mean value of --1.4 $\mathrm{km}\,\mathrm{s}^{-1}$, an rms of 11.1 $\mathrm{km}\,\mathrm{s}^{-1}$, 
and a median absolute deviation of 9.5 $\mathrm{km}\,\mathrm{s}^{-1}$. The largest deviation is --92.1 $\mathrm{km}\,\mathrm{s}^{-1}$, however; there are only 
5 pixels that have deviations above the channel spacing. We show the \HI\ position-velocity diagrams from the TiRiFiC model and the observed data cubes in 
Fig.~\ref{fig:modvsobspvd}. Overall, they agree with each other. However, the position-velocity diagram from slice C 
shows the largest discrepancy between the model and the data. As mentioned before, this is mostly due to the central 
absorption that has been blanked and set to zero fluxes before performing the fit.

\subsection{Halo gas in NGC 4945}\label{sec:halogas}
\subsubsection{Properties of the halo gas}
\begin{figure*}
\begin{tabular}{cc}
\includegraphics[scale=0.56]{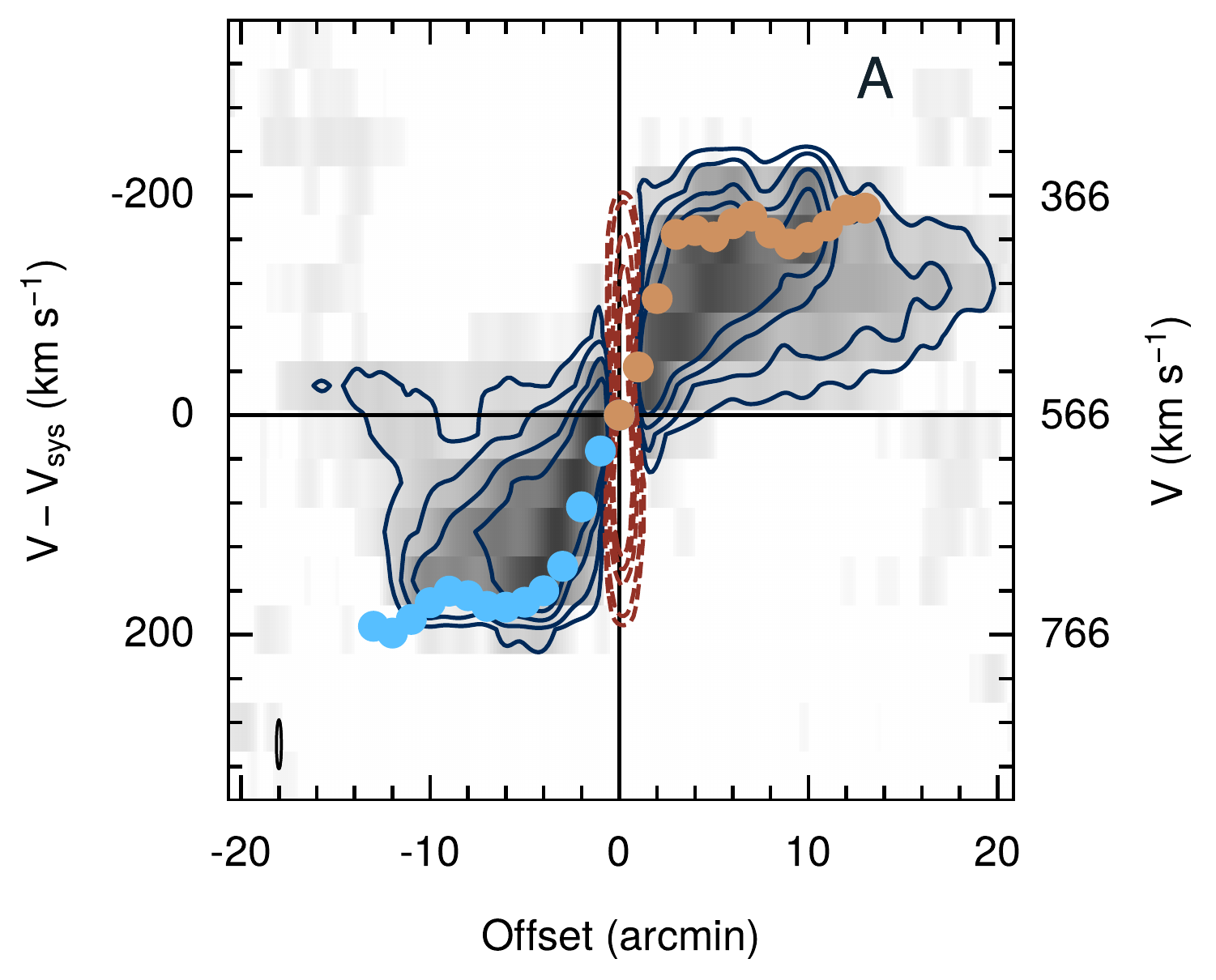}&
 \includegraphics[scale=0.56]{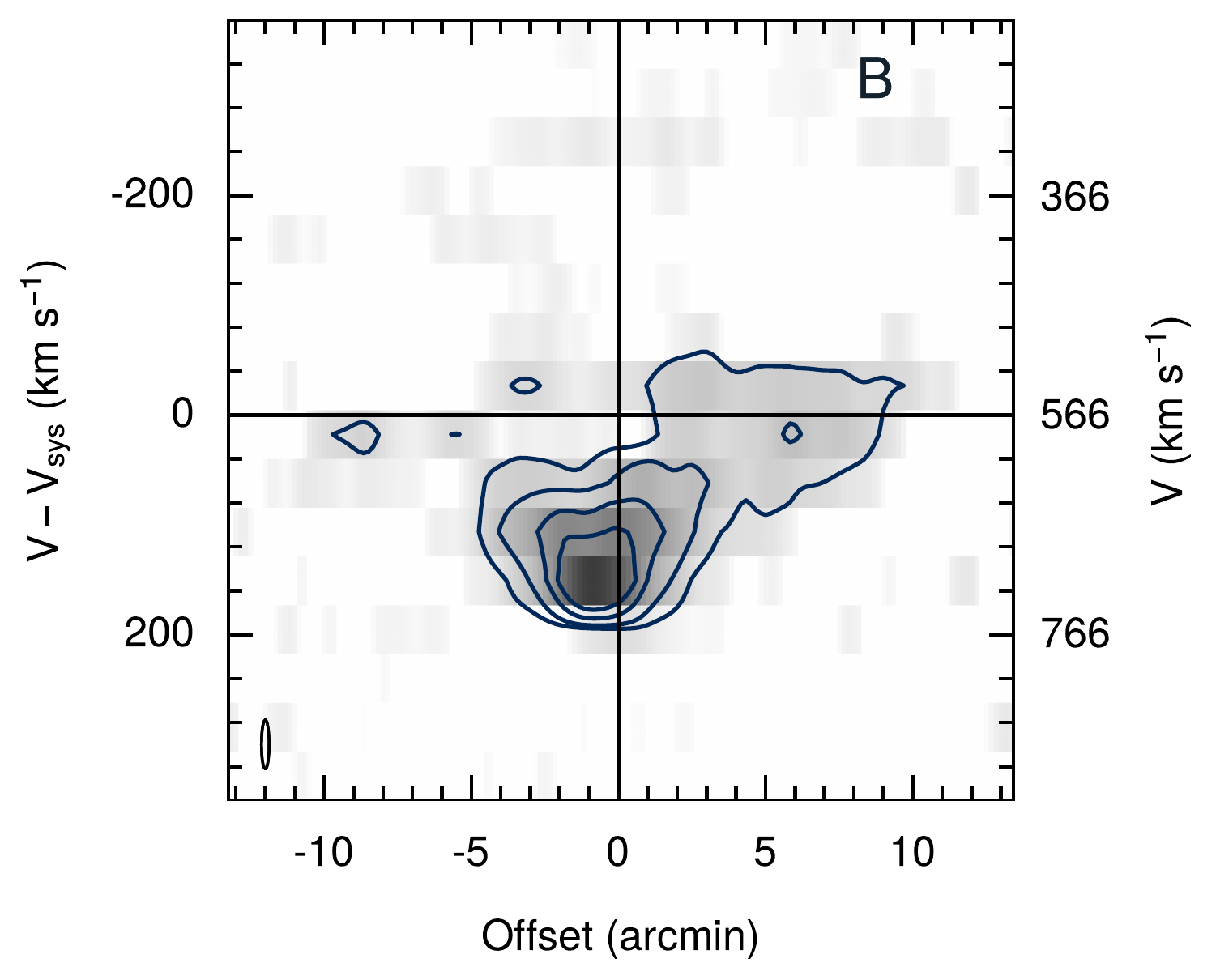}\\
 \includegraphics[scale=0.56]{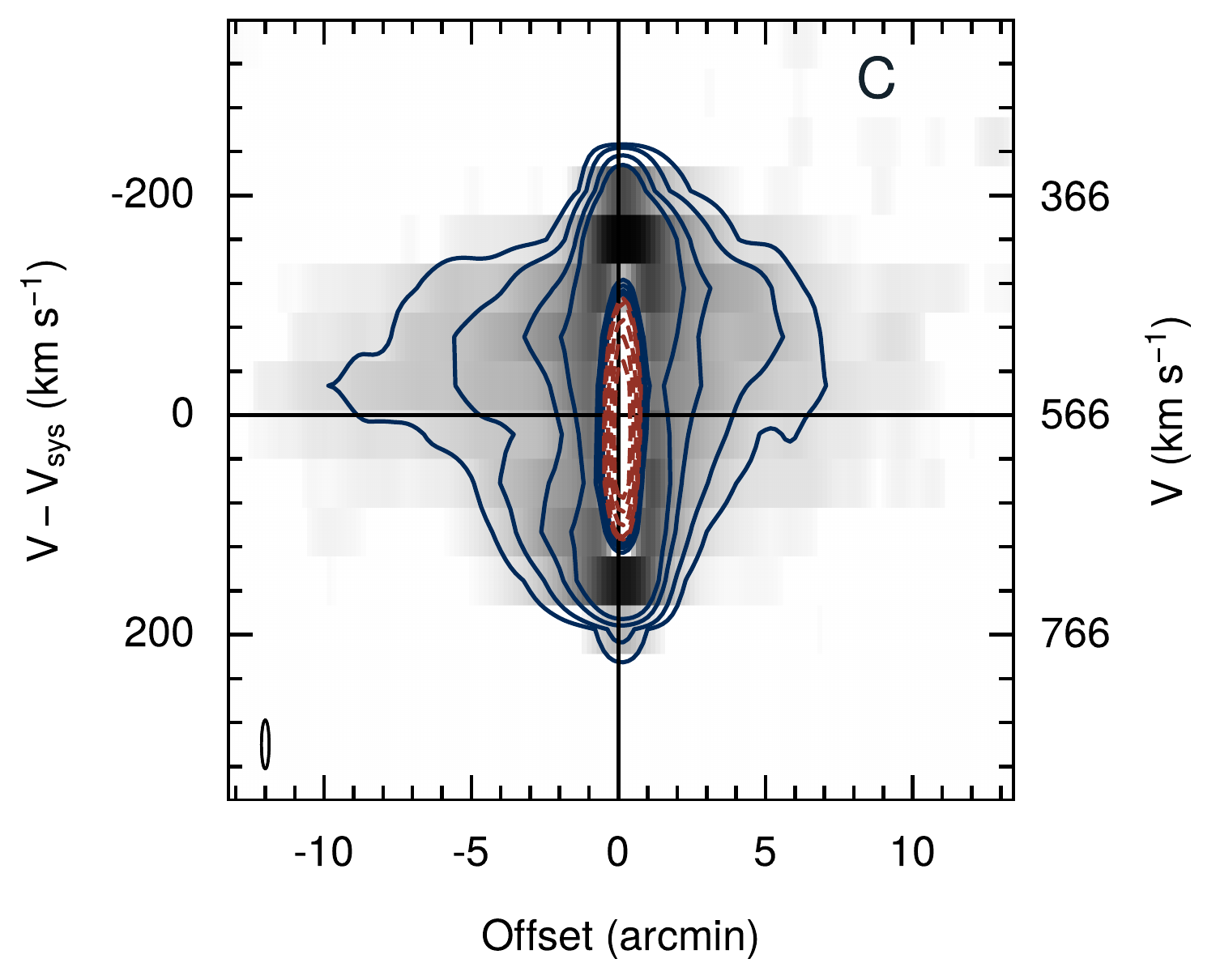}&
 \includegraphics[scale=0.56]{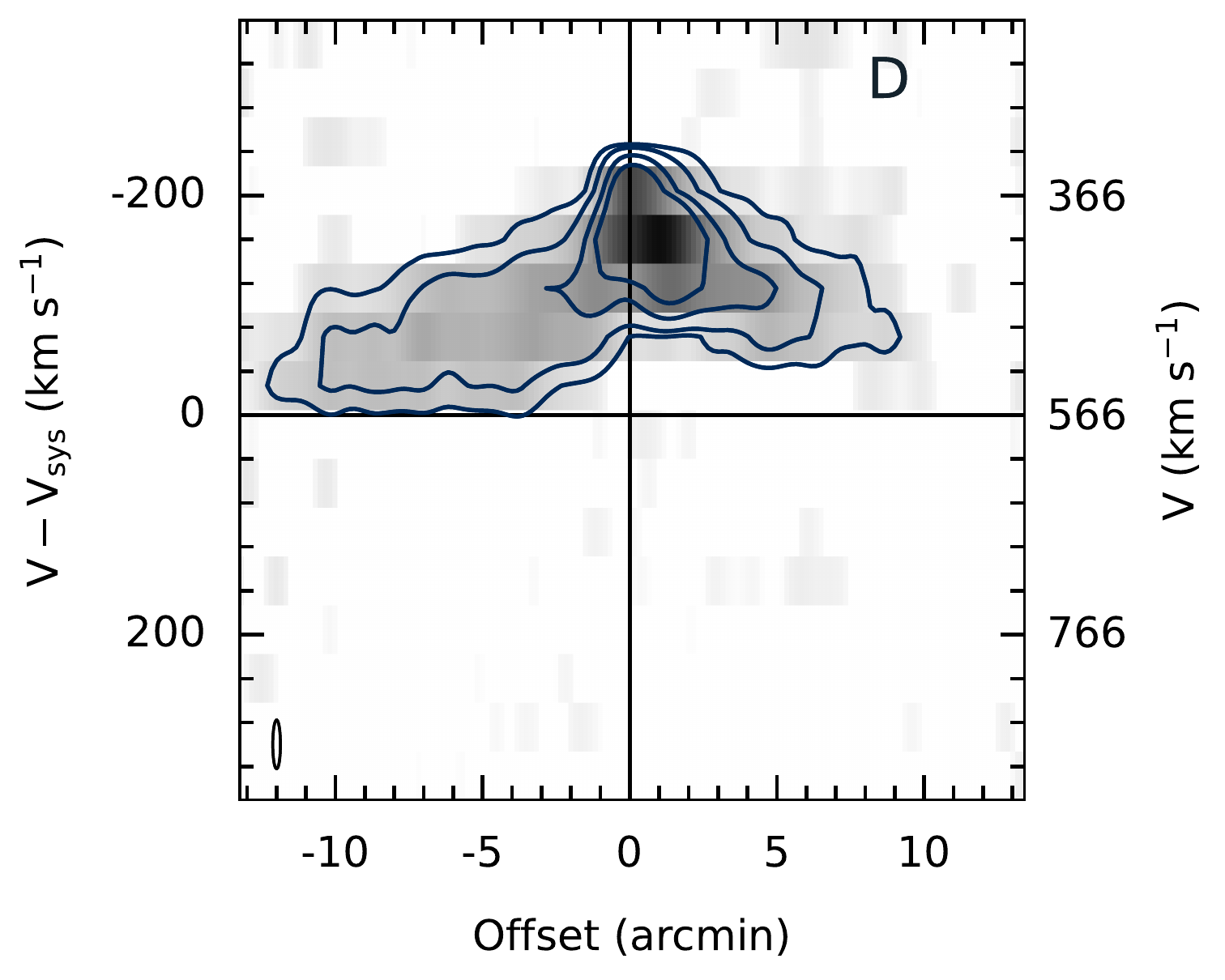} 
 \end{tabular}
 \caption{MeerKAT \HI\ position-velocity diagrams of NGC~4945 taken 
 along slices A, B, C, and D as shown in Fig.~\ref{fig:modvsobsmom}. Blue solid contours: \HI\ emission; the contour levels are (2.5, 6, 14, 24, 100) $\times$ rms, where rms is equal to 0.2 $\mathrm{mJy~beam^{-1}}$.  
 The circle symbols at the upper left panel show the TiRiFiC model rotation curve. Red dashed contours: \HI\ absorption; the contour levels are (-14.0 , -10.0 ,  -6.0 ,  -2.0,  -0.8) $\mathrm{mJy~beam^{-1}}$.
 The elongated ellipses at the lower left corners of the plots show the size of the beam FWHM = \SI{60}{\arcsecond} (along the x-axis) and the channel spacing (along the y-axis).}
 \label{fig:modvsobspvd1}
\end{figure*}

As mentioned previously, our high-sensitivity observations have revealed the presence of previously unseen emission 
in the halo of NGC 4945. To highlight this, we show the position-velocity diagrams of NGC 4945 in Fig.~\ref{fig:modvsobspvd1}. This clearly 
reveals the presence of two components: a bright \HI\ disc with flat rotation curve, and a faint halo gas with velocities closer to systemic than the 
regularly rotating disc. As shown by all the position-velocity cuts, the halo gas has an asymmetric distribution. 
Halo gas with similar properties to the one we find for NGC 4945 has been reported in both normal and starburst galaxies 
\citep{2001ApJ...562L..47F, 2015MNRAS.450.3935L, 2017ApJ...839..118V}. We will discuss the possible origin of the extra-planar gas in NGC 4945 
in Section~\ref{sec:discussion}.
\subsubsection{Separating the halo gas from the disc}
\begin{figure*} 
\centering
\includegraphics[scale=0.82]{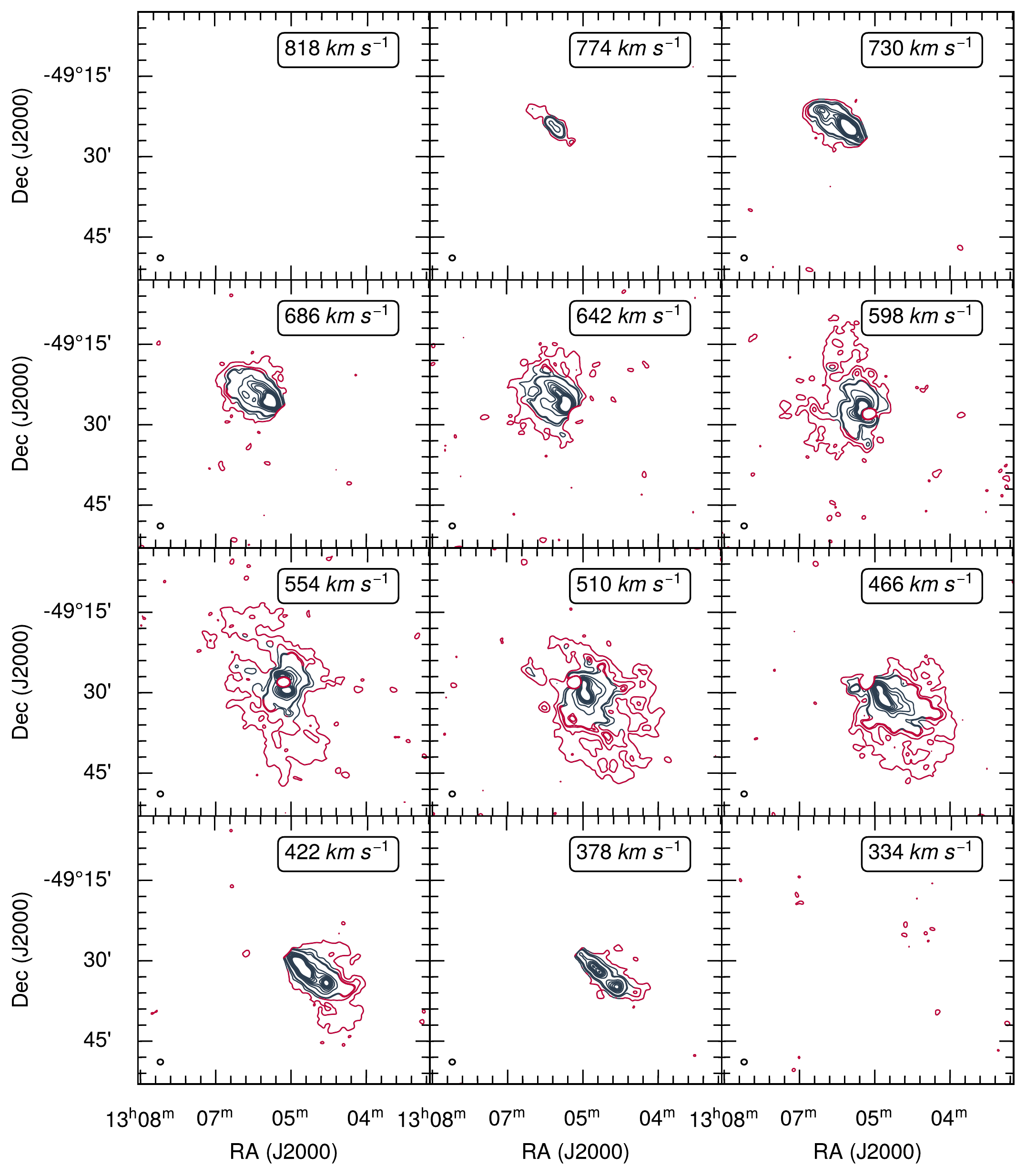}
\caption{MeerKAT H\,{\sc i} channel maps of the nearby starburst galaxy NGC~4945 showing the location of the anomalous gas (red contours) and the gas in the main 
disc of the galaxy (blue contours). 
The contour levels range from 3 $\times$ rms to 20 $\times$ rms in step of 7 $\times$ rms and 20 $\times$ rms to 400 $\times$ rms in step of 20 $\times$ rms where rms is 0.20 mJy\,beam$^{-1}$. The black 
circles shown at the bottom left corner of each panel show the beam (FWHM = \SI{60}{\arcsecond}).}
\label{fig:anom}
\end{figure*}

Due to our limited velocity resolution, the individual \HI\ profiles of NGC 4945 could not be decomposed into multiple Gaussian components as was done for NGC 253 by \citet{2015MNRAS.450.3935L}.
To summarize, they visually inspected each PV slice aligned along the galaxy's major axis to spot any kinematically anomalous gas 
component. In addition, they interactively fitted three Gaussians to all the line profiles making each PV slice. In our case, even using two Gaussians to fit the profiles failed in most cases. 
Thus, to separate the anomalous \HI\ from the galaxy's disc, we use the kinematic model data cube derived by TIRIFIC as outlined above to mask the main disc of the galaxy and isolate the kinematically anomalous gas. 
We show the location of the anomalous gas in Fig~\ref{fig:anom}. We tried to derive the rotation curve of the anomalous gas separately but we failed to obtain reliable kinematics.

We found an \HI\ mass of \num{3.7d8} $\mathrm{M_{\odot}}$, which accounts for 6.8\% of the total \HI\ mass of the galaxy. This is almost twice as large as the fraction of the anomalous gas 
found in NGC 253 \citep{2015MNRAS.450.3935L}. However, despite being a starburst galaxy, the fraction of the anomalous component we find for NGC 4945 is 
still well within the range expected for normal disc galaxies.  \citet{2019A&A...631A..50M} performed a Bayesian Markov chain Monte Carlo (MCMC) kinematic fit to 11 normal disc galaxies from the 
HALOGAS survey and found that about 5\% to 25\% of the \HI\ gas in their sample reside in the halo of the galaxies. 
Here, we expected to see more extra-planar \HI\ in NGC 4945 than in normal galaxies. This is because, 
as reported in the literature \citep[e.g.,][]{2000MNRAS.314..511S, 2000ApJS..129..493H}, 
the brief burst of intense star formation leads to the existence of superwind and 
multiple episodes of type II supernova explosions in starburst galaxies. The combined effects of these mechanisms are 
expected to drive more gas in the halo than seen in normal galaxies. However, as demonstrated by \citet{2000ApJS..129..493H, 2003RMxAC..17...47H}, 
starburst-driven superwind could accelerate a cloud of gas to a velocity larger than the velocity needed to escape the gravitational potential 
of the host galaxy ($v_{esc}$). This would result in a fraction of gas being driven to the surrounding intergalactic medium instead of ending up in the halo. 
\citet{2000ApJS..129..493H, 2003RMxAC..17...47H} demonstrated that the superwind's energy loss due to radiative cooling in starburst galaxies is not strong enough to halt this process. 
This may be why being a starburst galaxy does not simply translate to having more extra-planar \HI\ than normal galaxies. The escaped gas could be in the ionized form or a 
tenuous gas, escaping the detection limit of the current emission line surveys.

It is not uncommon to find more halo gas on one side than on the other side of galaxies, and this is true for both normal and starburst galaxies.
If we cut NGC 4945 into two halves, 20\% of the anomalous gas resides in the northern side of the minor axis, 
whereas 79\% of the halo gas resides in the southern side. Similarly, 42.4\% of the halo gas is located above the major axis, and 
57.6\% below the major axis of the galaxy. We have assumed a position angle of 45\degree for the calculation.

\begin{figure*} 
\begin{tabular}{l}
\centering
\includegraphics[scale=0.62]{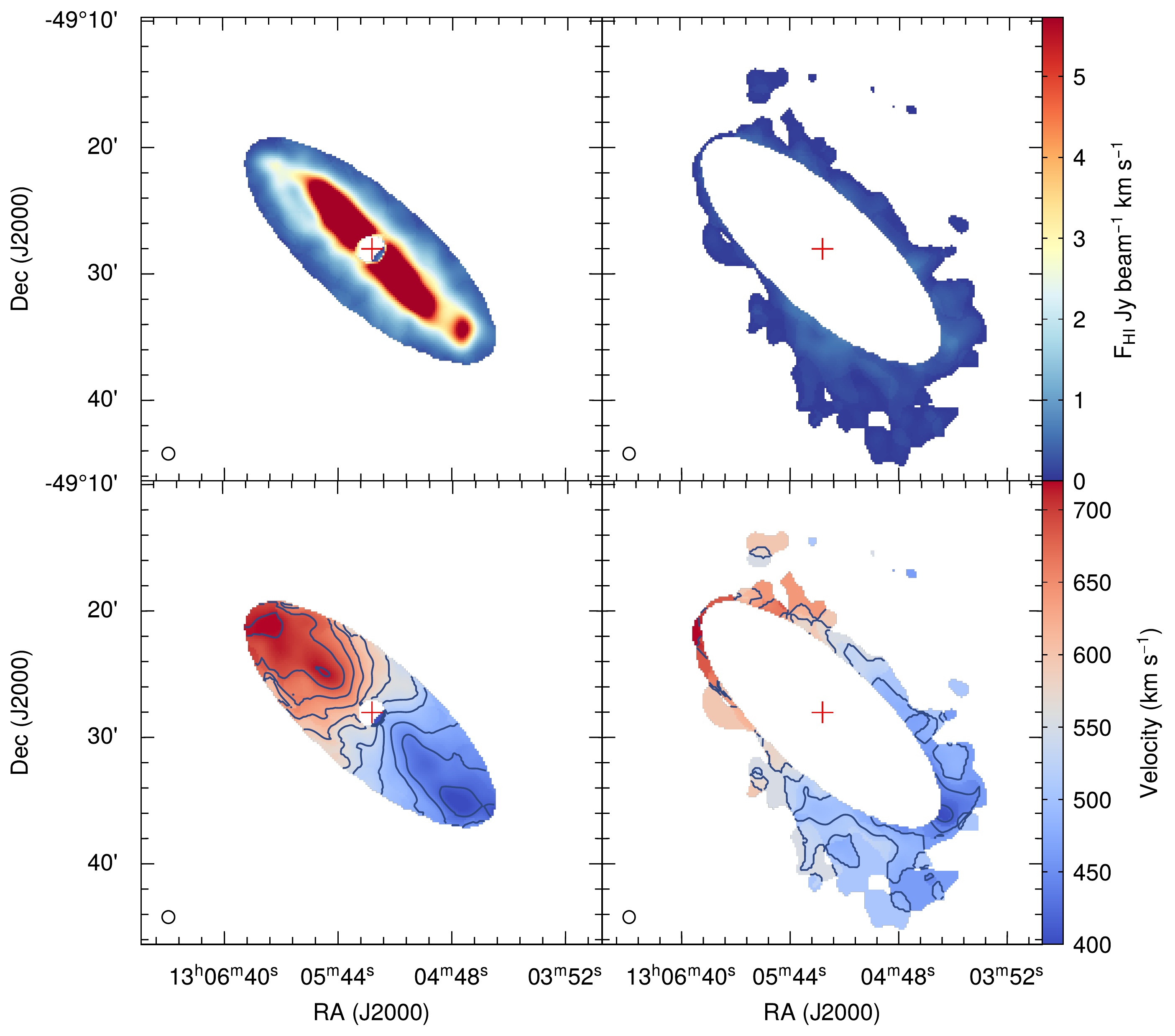}
\end{tabular}
\caption{Top left: Moment-0 map of the main disc of NGC 4945. Top right: moment-zero map of the anomalous gas in NGC 4945. Bottom left: 
velocity field (moment-1) of the main disc of NGC 4945. Bottom right: velocity field of the anomalous gas in NGC 4945. 
The contour levels are 385 $\mathrm{km~s^{-1}}$ to 745 $\mathrm{km~s^{-1}}$ in 
step of 30 $\mathrm{km~s^{-1}}$. The red crosses show the kinematic centre derived from TiRiFiC. The circles at the bottom left corner of 
each plot show the beam  (FWHM = \SI{60}{\arcsecond}).}
\label{fig:anom_mom1}
\end{figure*}

We highlight the velocity field of the anomalous component in Figure~\ref{fig:anom_mom1}. The iso-velocity contours of the halo gas are irregular, and in general, 
do not follow the overall rotation pattern of the main disc. If the gas is of a galactic fountain origin, which is plausible as NGC 4945 is a starburst galaxy, the irregularity 
of the velocity pattern of the halo gas indicates unsettled \HI\ that will be accreted back onto the main disc of the galaxy at a later stage. An outer disc star formation 
could also cause such an irregularity. This can be investigated further using deeper H$_{\alpha}$ imaging. However, the star formation in NGC 4945 is so far known to be confined in its central part.

\subsection{Radio continuum map}\label{sec:cont}
\begin{figure*} 
\begin{tabular}{l l}
\includegraphics[scale=0.66]{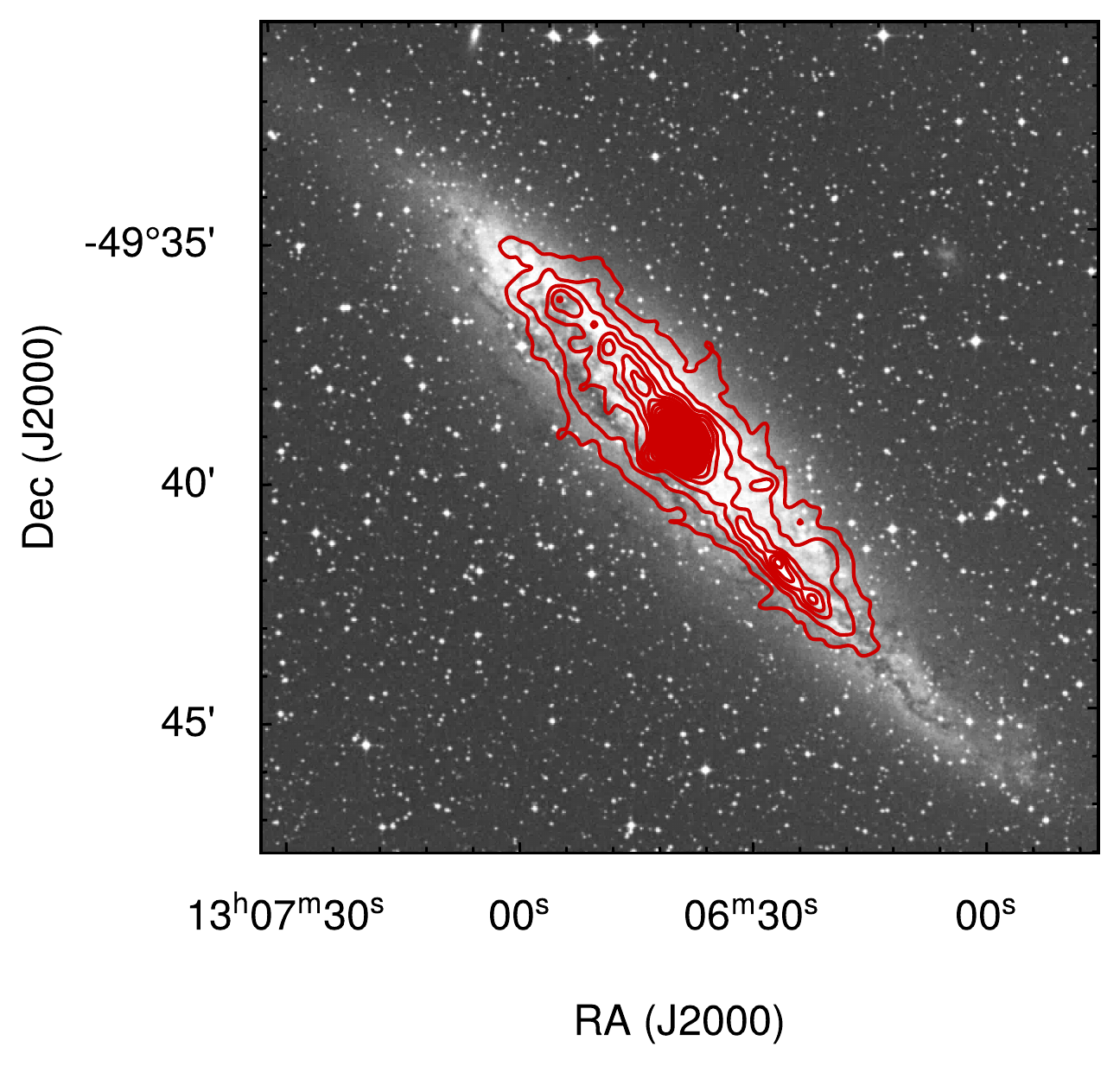}& 
\hspace*{-1.5cm}
\includegraphics[scale=0.66]{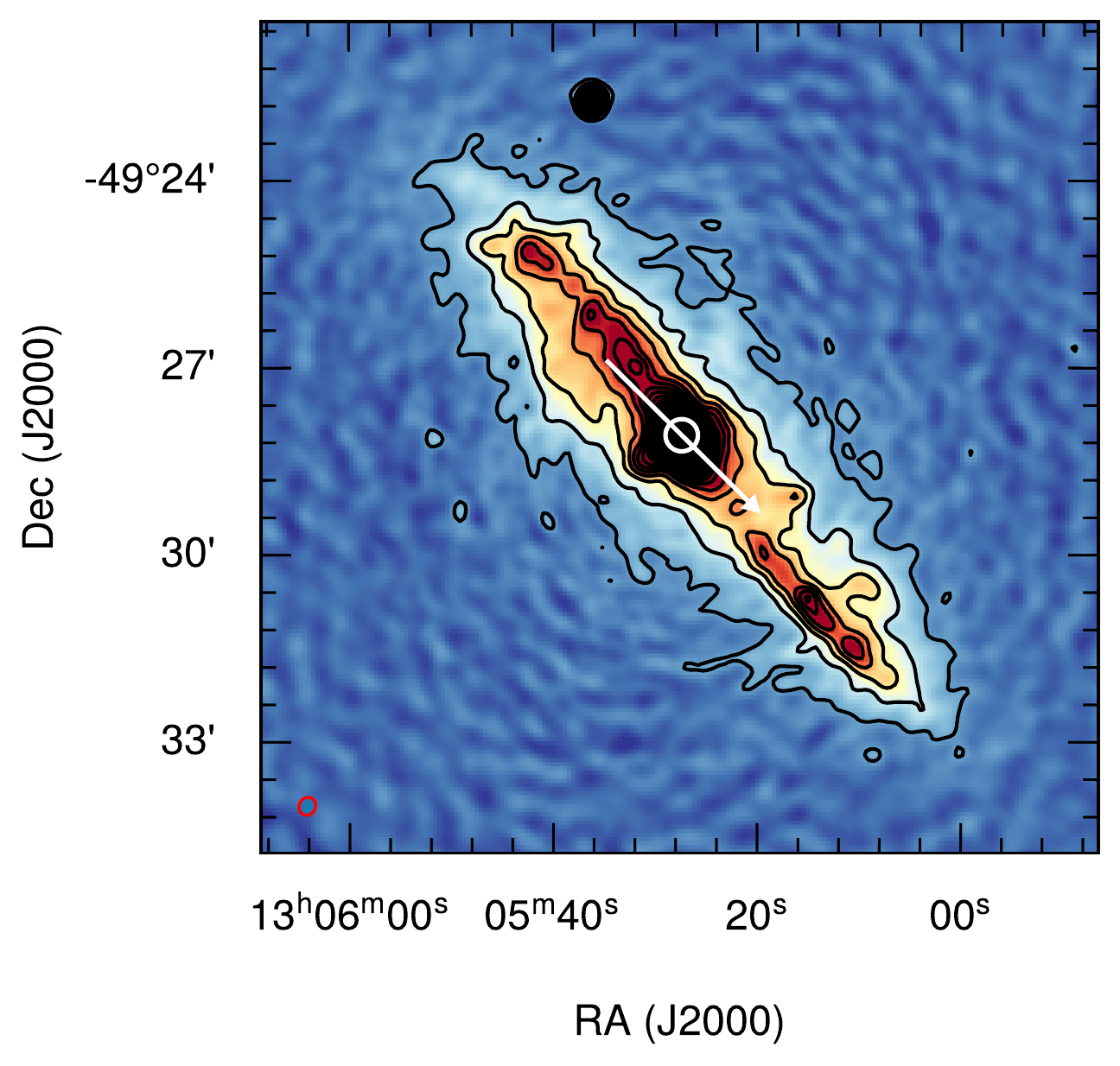}\\
\includegraphics[scale=0.606]{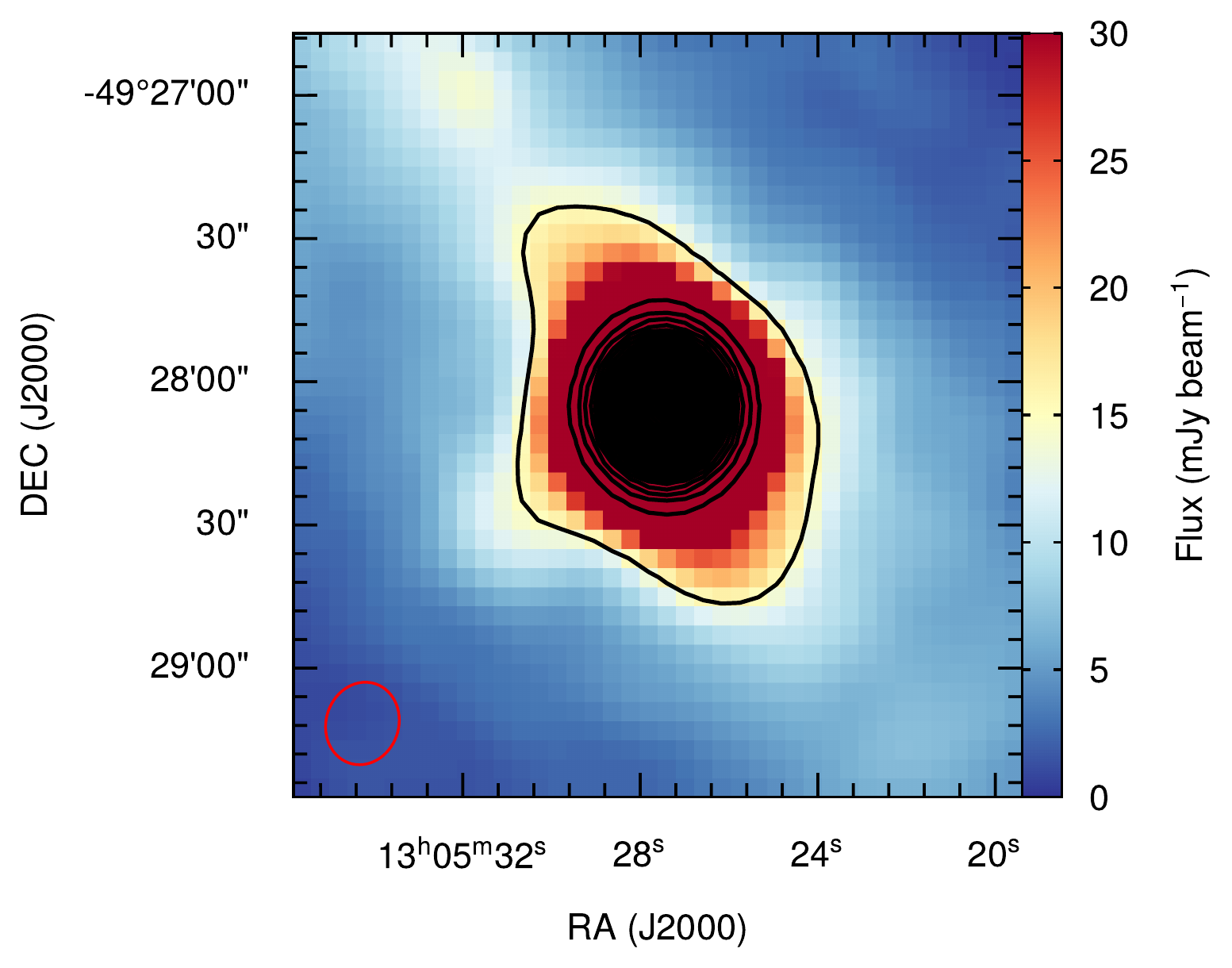}&
\hspace*{-0.5cm}
\includegraphics[scale=0.6]{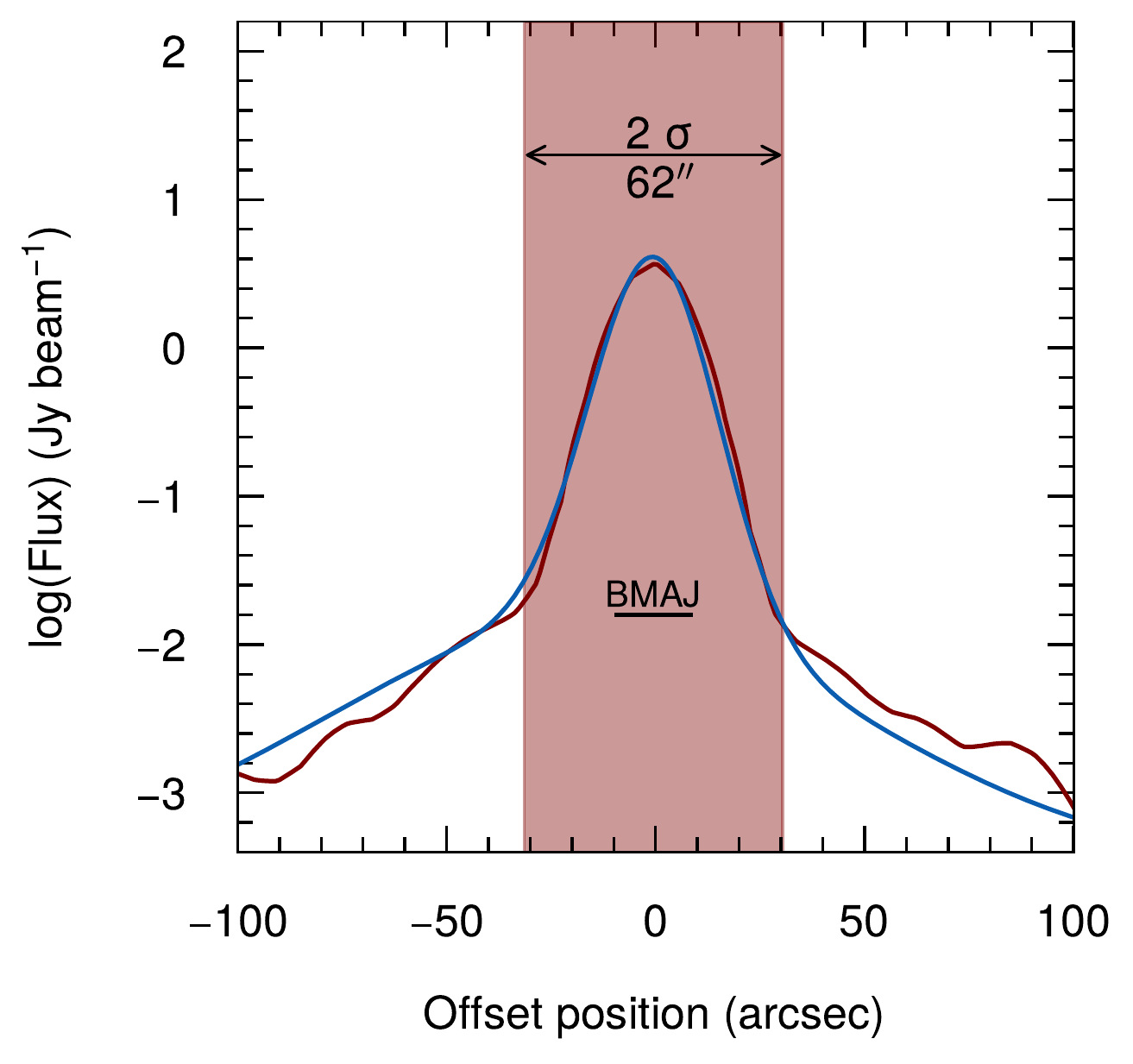}

\end{tabular}
\caption{Top left: MeerKAT radio continuum contour map of NGC~4945 (red contours) overlaid onto a DSS2 $B$-band optical image (grayscale). The 
contour levels are 0.00011 $\mathrm{Jy~beam^{-1}}$ to 3.8 $\mathrm{Jy~beam^{-1}}$ in step of 0.002 $\mathrm{Jy~beam^{-1}}$. 
Top right: continuum map of NGC~4945; the contour levels are 0.00011 $\mathrm{Jy~beam^{-1}}$ to 3.8 $\mathrm{Jy~beam^{-1}}$ in step of 0.002 $\mathrm{Jy~beam^{-1}}$. 
The red ellipse at the bottom left corner shows the beam (FWHM=\SI{17.6}{\arcsecond}). The white arrow, taken at a position angle of 45$\degree$, shows the slice from which the profile at the bottom left of the Figure 
was taken. The white circle shows the diameter of the core measured by fitting the profile at the bottom left of the Figure with a Gaussian function. The diameter of the core is \SI{31}{\arcsecond}. 
Bottom left: the central region of the continuum emission in NGC 4945. The contour levels go from 0.015 $\mathrm{Jy~beam^{-1}}$ to 3.72 $\mathrm{Jy~beam^{-1}}$ 
in step of 0.1 $\mathrm{mJy~beam^{-1}}$. The red ellipse at the bottom left corner shows the beam (FWHM=\SI{17.6}{\arcsecond}).  
Bottom right: flux density as a function of position along the slice shown as the white arrow at the top right panel of the Figure, plotted in log-scale to highlight the two discs 
components. The blue line shows a double Gaussian fit to the profile. The shaded area shows a 2-$\sigma$ width, where $\sigma$ is the dispersion of the (fitted Gaussian) 
core component. The value of $\sigma$ represents the diameter of the core of the disc shown by the white circle at the top right panel of the Figure. BMAJ represents the major axis of the beam.}
\label{fig:NGC4945-mom-cont}
\end{figure*}

\textbf{Distribution and total flux:} To produce the 20-cm radio continuum map of NGC~4945, we selected a 100 MHz chunk of the raw data around 
the central emission of NGC~4945 to be reduced by CARACal. WSClean was used in its multi-scale mode to account for the different size scales in 
the continuum map of NGC~4945. We let WSClean select the relevant scales automatically. 
Using a robust parameter of 0 and a Gaussian taper of FWHM = \SI{15}{\arcsecond}, the final beam size is 
\SI{17.6}{\arcsecond} $\times$ \SI{15.8}{\arcsecond}. 
The rms of the final residual continuum image is 0.053 $\mathrm{mJy~beam^{-1}}$, 
whereas the rms of the continuum image is 0.06 $\mathrm{mJy~beam^{-1}}$. 
We show an overlay of the radio continuum map of NGC~4945 onto a DSS $B$-band optical image in 
Fig.~\ref{fig:NGC4945-mom-cont}. The central continuum peaks at 
RA (J2000) = $\mathrm{13^{h}~05^{m}~27.4^{s}}$ and DEC (J2000) = \SI{-49}{\degree} \SI{28}{\arcminute} 
\SI{05}{\arcsecond}, which agrees well with the centre position found by 
\citet{1985PASAu...6..171W, 1997MNRAS.284..830E, 2001A&A...372..463O}. 
The emission extends over \SI{11.3}{\arcminute} $\times$ \SI{4.5}{\arcminute} 
at a position angle of \SI{45}{\degree} at a level of 1 $\mathrm{mJy~beam^{-1}}$. 
See the Appendix for an illustration on how these numbers were estimated. 
Using ATCA, \citet{2001A&A...372..463O} measured 
an extent \SI{11.6}{\arcminute} $\times$ \SI{3.3}{\arcminute} at a major axis position angle of \SI{45}{\degree}. Thus, we derive 
a similar major-axis extent as \citet{2001A&A...372..463O} but a slightly larger minor axis extent.\\

The continuum emission associated with 
the optical disc of the galaxy is shown in Fig.~\ref{fig:NGC4945-mom-cont}. 
The continuum is very extended and follows the optical disc of the galaxy, though its radial 
extent is much less than that of the optical disc.  It is characterised by a bright central core with a peak flux density of 
3.74 $\mathrm{Jy~beam^{-1}}$. In addition, there is a moderately bright, localised continuum emission along 
the major axis of the galaxy. Finally, a small extension of weak emission (which looks like an asymmetric double 
horned profile) is seen on the South-East side of the galaxy. 
Symmetric to that (with respect to the major axis), there seems to be a (finger-like) extension, although it is very weak. 
These may be signatures of outflows caused by star formation activity. We show a plot of the central continuum emission in 
Fig.~\ref{fig:NGC4945-mom-cont}. Our observation does not resolve the very compact nuclear core of NGC 4945. 
\citet{2001A&A...372..463O} found a source size of 
\SI{7.6}{\arcsecond} $\times$ \SI{3.4}{\arcsecond}. However, to get the central flux distribution at our resolution, 
we show a plot of the flux density as a function of 
position along a slice taken at a major axis position angle of \SI{45}{\degree} in Fig.~\ref{fig:NGC4945-mom-cont}. 
The profile has a bright narrow core and faint broad wings. We fit a double Gaussian function to the profile and derived
a dispersion of $\sigma$ = \SI{15.4}{\arcsecond}. As shown in Fig.~\ref{fig:NGC4945-mom-cont}, most of the 
emission in the core is contained within a diameter of twice this value, about \SI{31}{\arcsecond} (1 kpc at 6.7 Mpc).\\

To calculate the total continuum flux, first, we blank all continuum emission 
not associated with the galaxy using the MIRIAD task 
IMMASK. After that, we sum all the remaining non-blanked pixels and convert the unit from 
$\mathrm{Jy~beam^{-1}}$ to $\mathrm{Jy}$. We derive a total flux density of 6.2 Jy, of which 90\% resides 
within a radius of \SI{332}{\arcsec} (10.8 kpc). Thus, we recover $\sim$ 26\% more flux than \citet{2001A&A...372..463O}. A plot of the 
continuum map at the highest resolution (\SI{7.2}{\arcsecond} $\times$ \SI{6.3}{\arcsecond}) is shown in the Appendix. 
\\

\textbf{Star formation:} The radio continuum emission at 1.4 GHz can be used to trace recent star formation since it is dominated by synchrotron
radiation from supernovae remnants and thermal emission from H\,{\sc ii} regions \citep{2015ApJ...805...31L}. 
To convert the 21 cm continuum to a star formation rate (SFR), 
we first convert the continuum flux to luminosity using the following relation from \citet{2001ApJ...554..803Y, 2016A&A...585A..99M}:
\begin{equation}
 \log L_{1.4 \mathrm{GHz}}  \mathrm{[W~Hz^{-1}]} = 20.08 + 2 \log D \mathrm{[Mpc]} + \log S_{1.4 \mathrm{GHz}} \mathrm{[Jy]}, 
\end{equation}
where $D$ is the distance in Mpc, $L_{1.4 \mathrm{GHz}}$ is the radio continuum luminosity in $\mathrm{W~Hz^{-1}}$, and 
$S_{1.4 \mathrm{GHz}}$ is the total flux density in Jy. 
We derive $L_{1.4 \mathrm{GHz}}$ = \num{3.3e+22} $\mathrm{WHz^{-1}}$. 
The radio continuum luminosity is known to correlate with far-infrared (FIR) luminosity, indicating 
a strong correlation between star formation and cosmic ray production \citep[][and references therein]{2001ApJ...554..803Y}.
\citet{2001ApJ...554..803Y} analysed the FIR-radio continuum luminosity of 1809 IRAS 2 Jy sample galaxies taken from the 
1.2 Jy IRAS Redshift Survey catalogue \citep{1992ApJS...83...29S}, and found the 
following best-fitting correlation: 
\begin{equation}
    \log L_{1.4 \mathrm{GHz}} = (0.99 \pm 0.01) \log(L_{\SI{60}{\micro\metre}}/L_{\odot})~+~(12.07 \pm 0.08)
\end{equation}
NGC~4945 has a nuclear FIR luminosity of \num{4e+10} $L_{\odot}$. Using the above correlation, this corresponds to a 
1.4 GHz luminosity of $L_{1.4 \mathrm{GHz}}$ = \num{3.7e+22} $\mathrm{[W~Hz^{-1}]}$, 
which agrees with our observed value of 
$L_{1.4 \mathrm{GHz}}$ = \num{3.3e+22} $\mathrm{[W~Hz^{-1}]}$. We convert $L_{1.4 \mathrm{GHz}}$ to SFR 
 using the radio flux calibration of \citet{2003ApJ...586..794B} \citep[see also][]{2015ApJ...805...31L} given by: 
\begin{equation}
    \mathrm{SFR_{RC}} [M_{\odot} \mathrm{yr^{-1}}] = 
    \begin{cases}
      \num{5.52e-22} L_{1.4 \mathrm{GHz}} & \text{if $L_{1.4 \mathrm{GHz}} > L_{C}$ } \\
      \dfrac{\num{5.52e-22}}{0.1 + 0.9 (L/L_{C})^{0.3}} L_{1.4 \mathrm{GHz}} & \text{if $L_{1.4 \mathrm{GHz}} \leq L_{C}$}
    \end{cases} 
\end{equation}
, where $\mathrm{SFR_{RC}}$ is the radio continuum SFR in $M_{\odot} \mathrm{yr^{-1}}$, 
and $L_{C}$ = \num{6.4e+21} $\mathrm{W~Hz^{-1}}$ 
is the luminosity of a $\sim L{\ast}$ galaxy. Since $L_{1.4 \mathrm{GHz}}$ $\leq$ \num{e+23} for NGC~4945, 
the bulk of its radio continuum luminosity is expected to come from star formation rather than dominated by radio AGN 
\citep{2001ApJ...554..803Y}. We derive a star formation rate of $\mathrm{SFR_{RC}}$ = 18.4 $M_{\odot} \mathrm{yr^{-1}}$, 75\% of which comes 
from the core having a diameter of 1 kpc (defined previously as the area containing most of the bright continuum emission). 
This value is on the low-end of the range observed in starburst galaxies (10-100 $M_{\odot}$).

\section{Discussion}\label{sec:discussion}
This observation has changed our view of NGC~4945. We have found halo gas around almost the whole disc of the galaxies, not captured by previous \HI\ observations. 
For future high spectral resolution observations of this galaxy, it would be possible to do the interactive profile fitting techniques done by \citet{2015MNRAS.450.3935L} 
to separate the anomalous components in the galaxy and study the kinematics of the halo gas in detail. In this study, we have used the kinematic model of the galaxy to isolate 
the anomalous gas. Halo gas is found in virtually all sides of NGC 4945, with more \HI\ located in the lower-left quadrant of the galaxy. Halo gas with asymmetric distribution 
has previously been found for NGC~253. While \citet{2005A&A...431...65B} reported that the 
extra-planar gas in NGC~253 is only found in one side of the galaxy, the more sensitive map by \citet{2015MNRAS.450.3935L} showed that the halo gas is also found in the southern part of the galaxy. 
The distribution is still clearly asymmetric, with the northern side having more anomalous \HI\ than the south side. For the starburst galaxy NGC~253, 
it was shown that the gas in the halo of NGC~253 has similar kinematics to the gas 
in the disc but is trailing by 10--20 $\mathrm{km~s^{-1}}$ \citep{2015MNRAS.450.3935L}. Future high-resolution studies of NGC 4945 can confirm if this is also the case for NGC 4945.

Many scenarios are proposed in the literature regarding the origin of gas in the halo of galaxies. 
These include the interaction of galaxies with a nearby companion or the environment, pristine gas from the cosmic 
web accreted onto the galaxies, 
gas being blown out of the disc into the halo by expanding superbubbles created by supernovae and stellar winds 
\citep{2014A&ARv..22...71S, 2006MNRAS.366..449F, 2019A&A...631A..50M}.
NGC~4945 belongs to the Centaurus A group, and thus interaction with the intra-group medium may have played a role in forming its halo gas. 
The likelihood of the halo gas being 
from the cosmic web is low given the column density level at which we are detecting it. Accretion from the cosmic web is expected to happen at a lower column density level 
than we are probing. Thus, although we were unable to model the kinematics of the halo gas and compare to simulations, at least part of 
the halo gas we see here is likely due to 
outflows driven by the central starburst of the galaxy as seen in NGC 1808 and NGC 253 \citep{1993ApJ...402L..41K, 2015MNRAS.450.3935L}.  \\ 


The rotation curve of NGC~4945 resembles the overall spiral galaxy populations, most of which are characterised by a 
steep solid-body inner rotation curve and then flattens out to the outermost observed radius. 
There are exceptional cases where rotation curves are declining, such as that of NGC~253 \citep{2011MNRAS.411...71H, 2015MNRAS.450.3935L}, 
see also the cases for 22 nearby spiral galaxies compiled from the literature by \citet{2020ARep...64..295Z}. We did not find a clear signature of a declining rotation curve for NGC~4945. 
As explained in \citet{2020ARep...64..295Z}, cares need to be taken when interpreting a declining 
rotation curve since it does not necessarily indicate the edge of the dark matter halo.

We have fit the receding and the approaching sides of the galaxy separately to derive the rotation curve of NGC~4945. We 
have found that both sides have an outer warp. The warp on the receding side is more clearly visible in the models and the observed velocity field. 
The intensity distribution in the receding side appears to be suppressed since it extends just as much as the optical disc. 
This type of halo asymmetry has been found for NGC 253 \citep{2005A&A...431...65B}. 
They suggested that such a lack of \HI\ might be caused by ionization resulting from the galaxy's starburst activity. For this to occur, though, the outer layer of the 
gas needs to be warped, which is the case for NGC 4945. Tidal interactions and ram pressure 
stripping are among the mechanisms that 
can cause \HI\ deficiency. Perhaps the warp in the receding side is related to this \HI\ suppression. However, it is worth noting that \HI\ deficient galaxies are not 
necessarily warped. The origin of a warp is still a subject of intense investigation in the literature \citep{2006MNRAS.365..555S, 2020NatAs...4..590P}. 
Re-observation of this 
galaxy at higher velocity resolution will enable us to do proper modelling of the kinematics of the halo gas and investigate its possible origin based on existing 
models.  
\section{Summary}
We have presented MeerKAT \HI\ maps of the starburst galaxy NGC~4945, revealing for the first time the presence of faint \HI\ emission in its outskirts. 
The halo gas accounts for about 6.8\% of the galaxy's total mass. 
This is thanks to the sensitivity of MeerKAT, enabling us to go down to 
a $3\sigma$ column density level of \num{5e+18} $\mathrm{cm^{-2}}$. Despite being a starburst galaxy and the improved sensitivity, 
the amount of the halo gas in NGC 4945 is within the range found for normal star-forming galaxies. This could indicate that there might be gas expelled from the galaxy into 
the intergalactic medium due to a starburst-driven superwind  \citep{2000MNRAS.314..511S, 2000ApJS..129..493H}. 
In addition, the halo gas does not seem to follow the kinematics of the disc (such as the trailing gas in NGC 891 or NGC 253), 
this could be another indication that there might be gas escaping the galaxy. After isolating the area of genuine emission, we derived a 
total \HI\ flux density of 509 Jy\,km~s$^{-1}$, which is about 20\% higher than the flux density derived by \citet{2018MNRAS.478.1611K} 
using ATCA mosaic imaging. We have modelled the \HI\ emission through 3D tilted-ring modelling techniques using 
the TiRiFiC and FAT software \citep{2007AA...468..903J, 2015MNRAS.452.3139K}. The halo gas does not follow the regular kinematics of the main disc of the galaxy, 
and our attempt to fit it with TiRiFiC failed. Therefore, we have only made the kinematic modelling of the main disc of the galaxy. The modelling indicates the presence of 
a warp on the approaching and the receding sides of the galaxy. We derive a flat rotation curve with $v_{flat}$ = 
176 $\mathrm{km~s^{-1}}$ for the approaching side and $v_{flat}$ = 188 $\mathrm{km~s^{-1}}$ for the receding side, in agreement with \citet{2015MNRAS.452.3139K}. 
Due to our coarse velocity resolution, only areas of bright \HI\ emission were modelled here. 
Strong \HI\ absorption seen throughout the entire velocity range seen in \HI\ emission, is present towards the nuclear region of NGC~4945. The absorption 
lines appear to be broad and asymmetric, indicating the presence of a fast rotating ring and outflowing gas from the nuclei. 
The existence of a fast rotating ring is further supported by the clear rotation pattern in the absorption velocity field of NGC~4945. 
This fast rotation can be caused by the interaction of the surrounding neutral gas with the AGN at the centre of NGC~4945. 
A similar case has been found for NG~1808 by \citet{1993ApJ...402L..41K}.
We have also presented a map of the radio continuum emission, most of which is associated with the bright optical disc of NGC~4945. 
The continuum emission is very extended and characterised by a bright central core with a diameter of about 0.6 kpc. 
The continuum flux suggests that NGC~4945 is forming stars at a global rate of 18.4 $M_{\odot} \mathrm{yr^{-1}}$, 75\% of which resides within 
a core radius of 1 kpc.  
\section{Data availability}
The data from this study are available upon request to the corresponding author, Roger Ianjamasimanana.
\section{Acknowledgements}
We would like to thank the anonymous referee for a careful reading of the manuscript and very useful comments, which greatly improved 
the presentation of the paper. \\\\
RI acknowledges financial support from grant RTI2018-096228-B-C31 (MCIU/AEI/FEDER,UE) and from the State Agency for Research of the Spanish Ministry of Science, Innovation and Universities through the "Center of Excellence Severo Ochoa" awarded to the Instituto de Astrofísica de Andalucía (SEV-2017-0709), from the grant IAA4SKA (Ref. R18-RT-3082) from the Economic Transformation, Industry, Knowledge and Universities Council of the Regional Government of Andalusia and the European Regional Development Fund from the European Union. \\\\
The MeerKAT telescope is operated by the South African Radio
Astronomy Observatory,  which is a facility of the National Research
Foundation, an agency of the Department of Science and Innovation.\\\\
This work is based upon research supported by the South African
Research Chairs Initiative of the Department of Science and Technology
and National Research Foundation.\\\\
The financial assistance of the South African Radio Astronomy Observatory (SARAO) towards this research is hereby acknowledged (\url{www.sarao.ac.za}).\\\\
At Ruhr University Bochum this research is supported by BMBF Verbundforschung grant 05A20PC4 and by DFG Sonderforschungsbereich SFB1491.
\\\\
This project has received funding from the European Research Council (ERC) under the European Union’s Horizon 2020 research and innovation programme grant agreement no. 882793, project name MeerGas.
\\\\
This project has received funding from the European Research Council (ERC) under the European Union's Horizon 2020
research and innovation programme (grant agreement no. 679627; project name FORNAX). 
\bibliographystyle{mnras}
\bibliography{biblio_mnras} 



\appendix
\section{Additional Figures}
\begin{figure*}
\begin{tabular}{c}
 \includegraphics[scale=0.7]{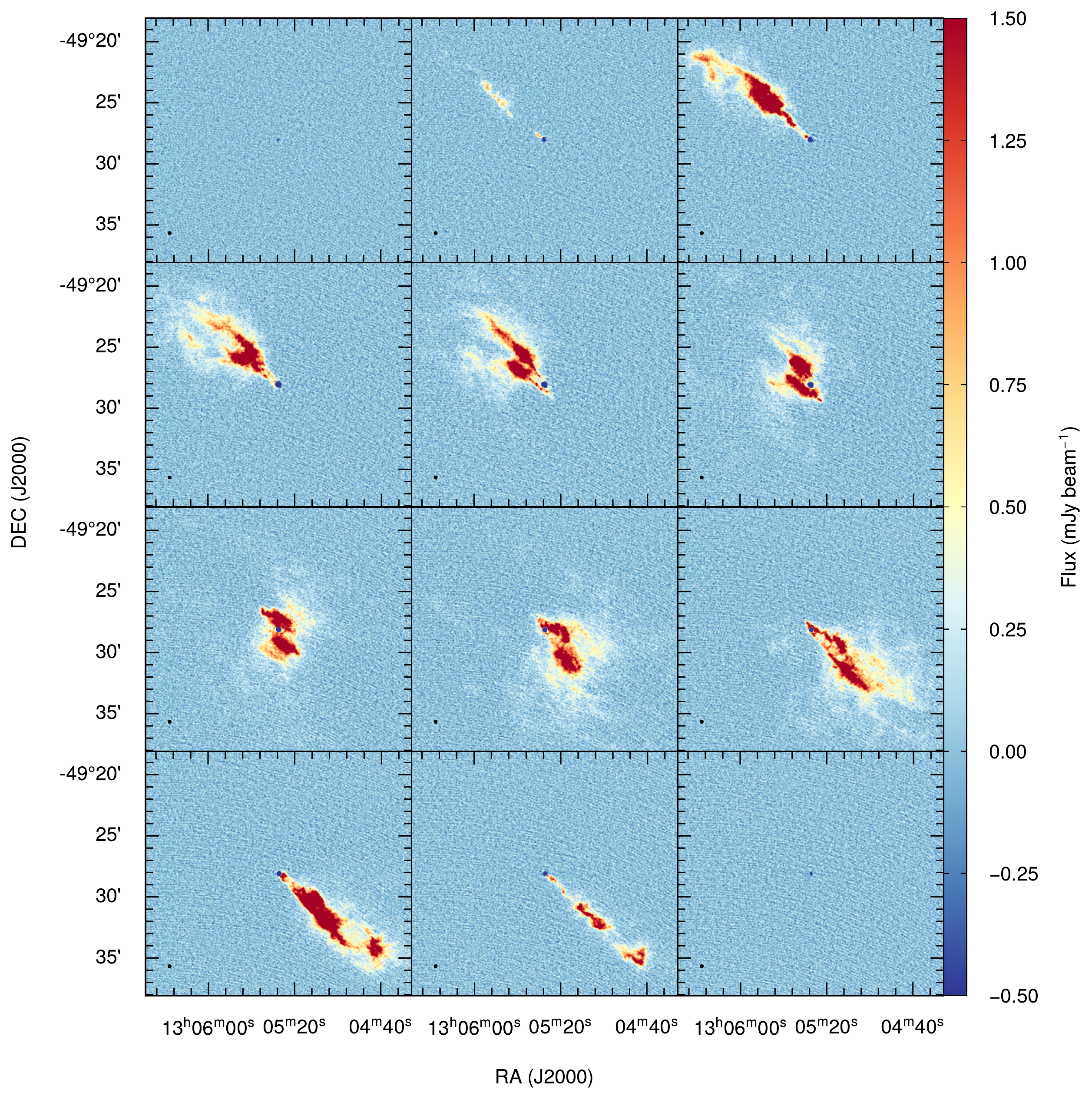}
 \end{tabular}
 \caption{MeerKAT \HI\ channel maps of NGC 4945 at a resolution of \SI{7.5}{\arcsecond} $\times$ \SI{6.4}{\arcsecond}. The size of the beam is represented 
 by the small dots at the bottom left corner of each panel.}
 \label{figap:channel}
\end{figure*}

\begin{figure*}
\begin{tabular}{cc}
    \includegraphics[scale=0.63]{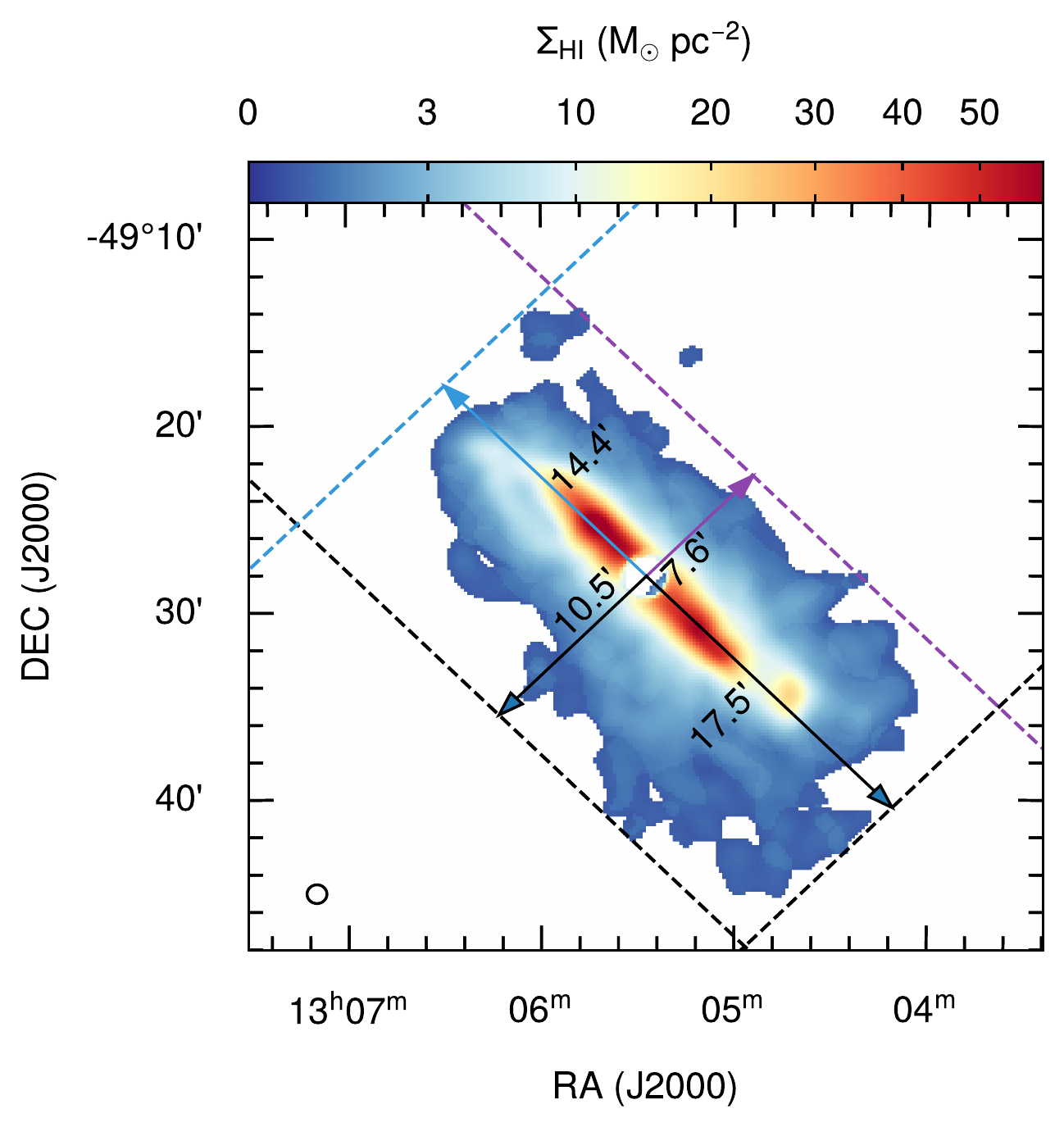}&
    \includegraphics[scale=0.63]{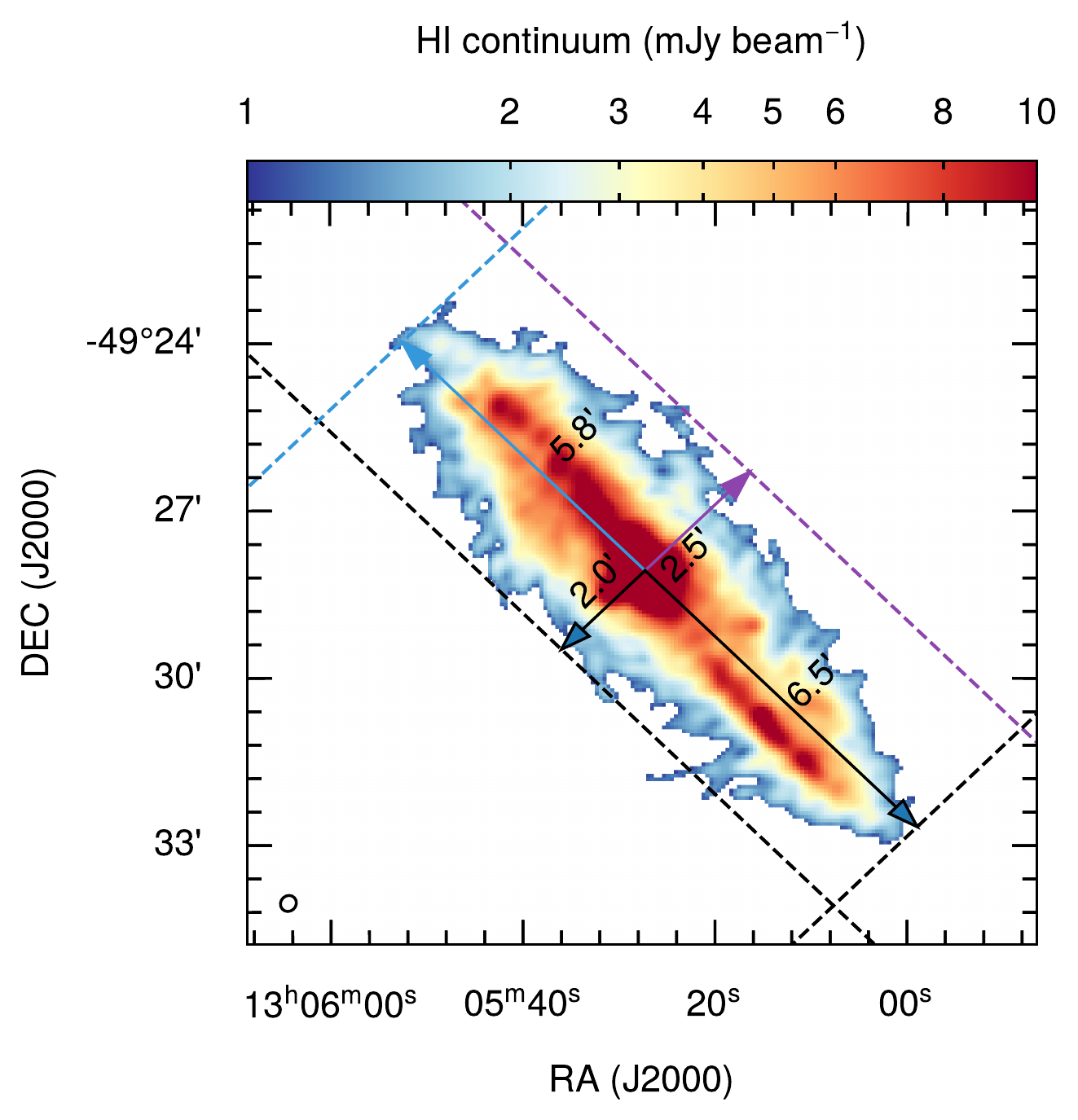}\\
    \includegraphics[scale=0.63]{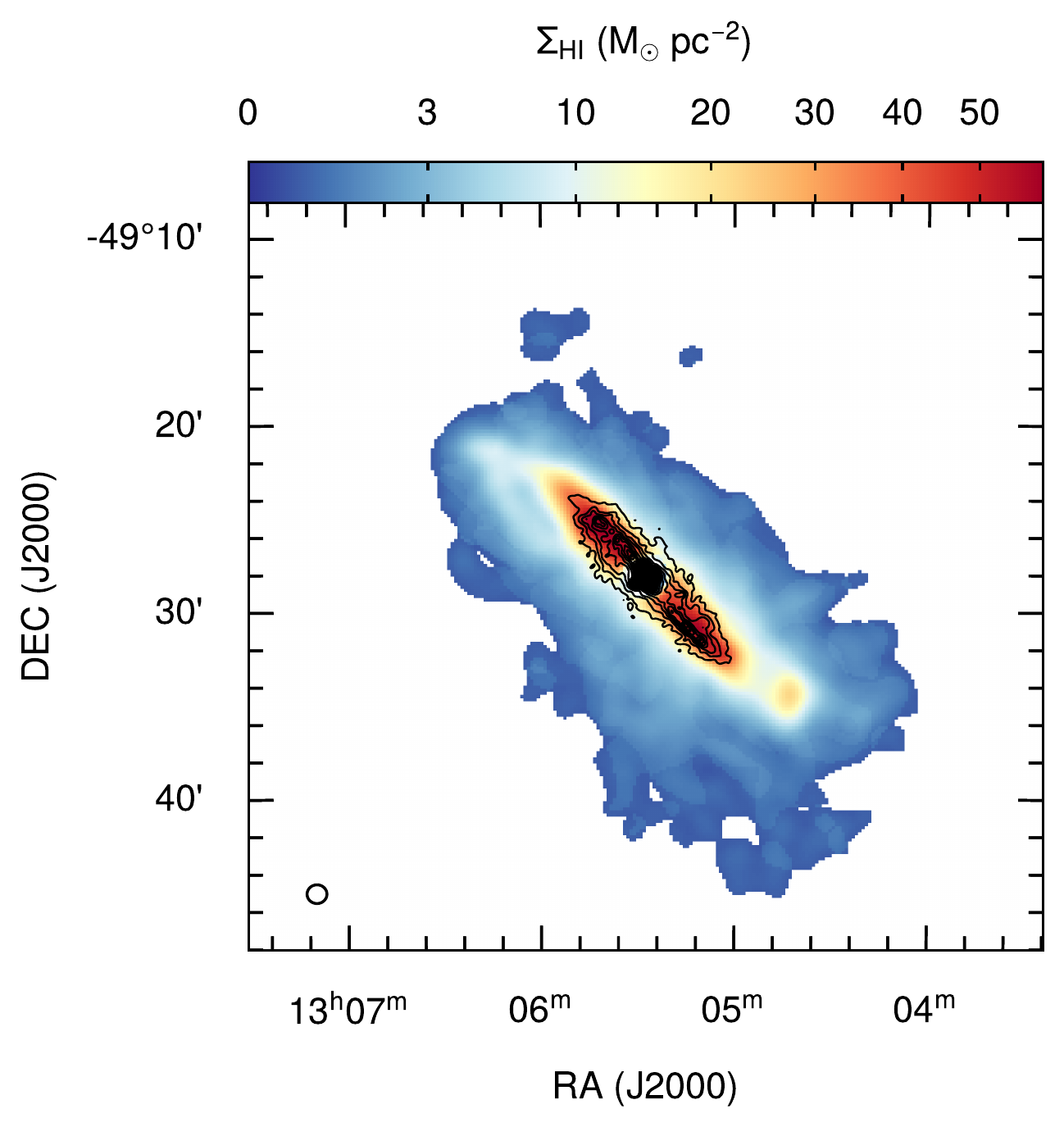}&
    \includegraphics[scale=0.63]{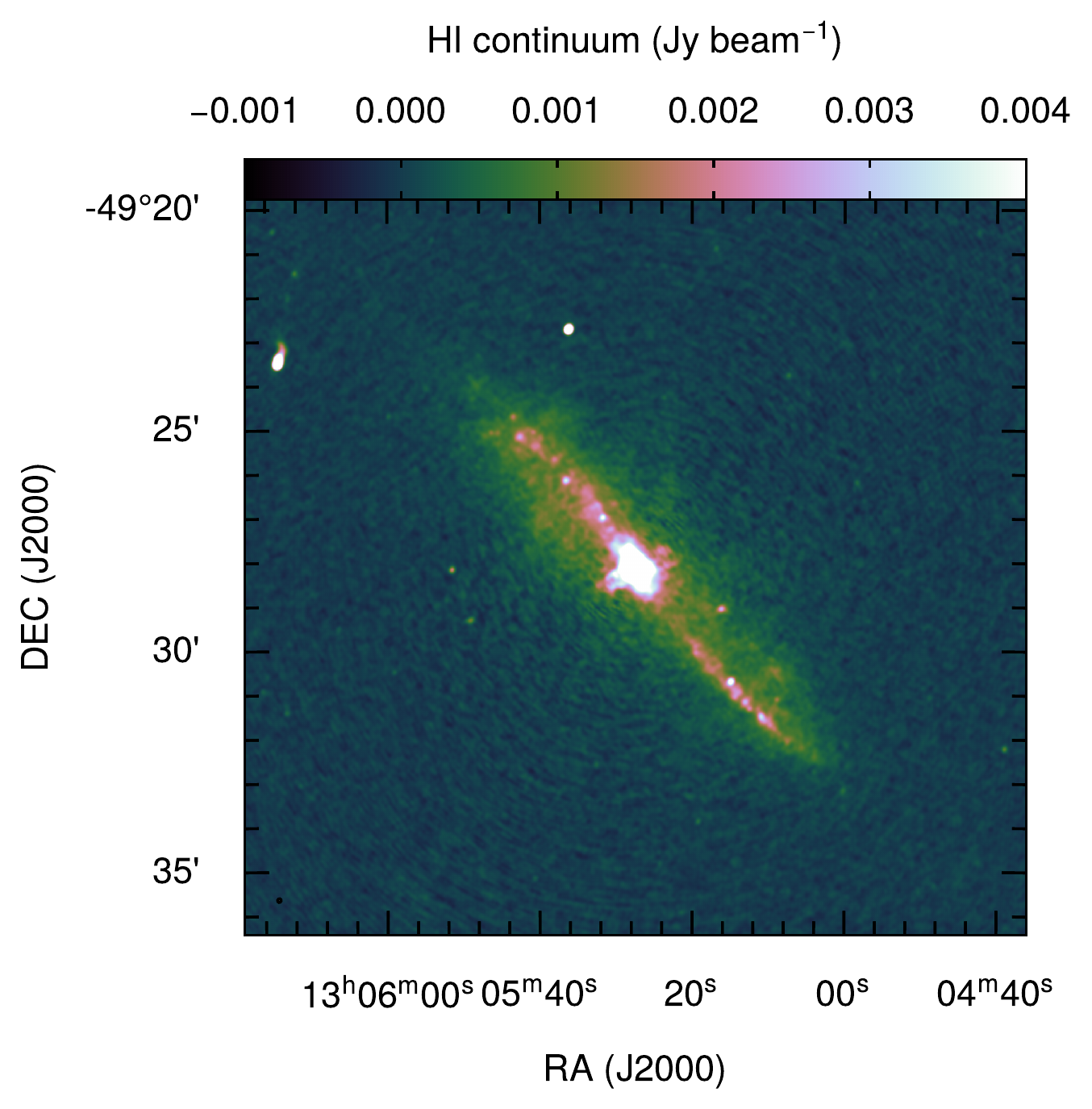}
  \end{tabular}
  \caption{
 Top left: \HI\ surface density map of NGC~4945, arrows and dashed lines indicate the extent of the \HI\ disc. The beam is shown by the black circle 
 at the bottom left corner of the plot (FWHM = \SI{60}{\arcsecond}) .
 Top right: continuum map of NGC~4945, arrows and dashed lines indicate the extent of the continuum emission. The beam is shown by the black ellipse 
 at the bottom left corner of the plot ($\mathrm{FWHM_{BMAJ}~\times~FWHM_{BMIN}}$ = \SI{17.6}{\arcsecond} $\times$ \SI{15.8}{\arcsecond}). 
 Bottom left: \HI\ surface density map (coloured image) and the continuum map of NGC 4945 (black contours). The contour levels range from 0.0001 $\mathrm{Jy~beam^{-1}}$ to 3 $\mathrm{Jy~beam^{-1}}$ 
 in step of 0.0015 $\mathrm{Jy~beam^{-1}}$. The black circle at the bottom left corner of the plot shows the beam (FWHM = \SI{60}{\arcsecond}). 
 Bottom right: the radio continuum map of the galaxy at a resolution of \SI{7.3}{\arcsecond} $\times$ \SI{6.2}{\arcsecond}. The beam is 
 shown at the bottom left corner of the plot but is very small so it is not visible.}
  \label{figap:continuum}
 \end{figure*}

In this Appendix, we show additional figures that complement the main section of the paper. The channel maps of NGC 4945 from the high-resolution data cube 
(\SI{7.5}{\arcsecond} $\times$ \SI{6.4}{\arcsecond}) is shown in Fig.~\ref{figap:channel}. As mentioned previously, even at this resolution, the fluffy faint \HI\ emission 
is already visible. The top left panel of Fig.~\ref{figap:continuum}, illustrates how we determine the extent of the \HI\ disc as explained in Section~\ref{sub:moment} 
of the paper. Similarly, the top right panel of Fig.~\ref{figap:continuum} shows an illustration of the estimation of the extent of the continuum map 
as mentioned in Section~\ref{sec:cont}. To show the size of the continuum map relative to the size of the \HI\ disc of NGC 4945, we overlay the continuum map 
on top of its \HI\ surface density map. Finally, a high resolution version of the continuum map, i.e., at a resolution of~\SI{7.3}{\arcsecond} $\times$ \SI{6.2}{\arcsecond} 
is shown at the bottom right of Fig.~\ref{figap:continuum}. 

\bsp	
\label{lastpage}
\end{document}